\documentclass[
  aps,prx,twocolumn,
  superscriptaddress,
  longbibliography
]{revtex4-2}
\usepackage{empheq}
\usepackage{amsmath,amssymb,amsfonts}
\usepackage{amsthm}
\usepackage{graphicx}
\usepackage{bm}
\usepackage{hyperref}
\usepackage{tikz}

\begin{document}

\title{TCL4 Asymptotic Redundancy and \\ Canonically Consistent Master Equations}

\author{Dragomir Davidovic}
\email{dragomir.davidovic@physics.gatech.edu}
\affiliation{School of Physics, Georgia Institute of Technology, USA}

\date{} 

\begin{abstract}
Left alone, open quantum systems relax toward a Kubo--Martin--Schwinger (KMS) equilibrium state, yet the inner workings of this process remain opaque. It remains unclear why the intricate fourth-order time-convolutionless (TCL4) population generator reproduces the comparatively simple second-order stationary state corrections. Even more surprisingly, stationary-state corrections of such precision can be compressed into a simple virtual coherence pathway: an open quantum systems analogue of virtual transitions, in which populations communicate through coherences that are never occupied.
 Here we show that this simplification arises through a sequence of stationary-state-preserving transformations that progressively eliminate asymptotically redundant components while preserving the stationary state, ultimately yielding the virtual coherence pathway. This resolves the longstanding stationary-state problem of the Redfield equation and reveals that much of the apparent complexity of the TCL4 generator is asymptotically redundant.
\end{abstract}


\maketitle

\section{Introduction}
Open quantum systems are generally expected to relax toward a time-invariant Kubo--Martin--Schwinger state~\cite{davies1974,trushechkin2022open}. Just as the fluctuation--dissipation theorem characterizes equilibrium response, return to equilibrium (R2E) characterizes the dynamics approaching that state. This provides a reference process for reduced descriptions across quantum optics, condensed matter physics, nonequilibrium statistical mechanics, and quantum information theory. As a benchmark with a well-defined stationary state, it offers a rare opportunity to investigate how finite-coupling stationary-state information is encoded within the reduced dynamics.

Exact stationary states  have been established for a broad class of Gaussian open systems, including quantum diffusion models~\cite{grabert1988quantum,fleming2011exact,
subacsi2012equilibrium,AbbruzzoGiovannettiCavinaPRL2025}. In these
settings the reduced dynamics can be solved exactly.
In contrast, the stationary-state problem for generic finite-dimensional systems such as spin-boson and multilevel models remains much less understood, as exact solutions are generally unavailable. The only broadly established asymptotic
description is the ultra-weak-coupling limit, in which
the reduced dynamics converges to Davies generators and
the stationary state reduces to the Gibbs
state~\cite{davies1974,trushechkin2022open}.

Beyond the ultra-weak-coupling limit, finite-coupling stationary states have been studied using spectral and resonance methods~\cite{bach2000return,JaksicPillet2002,frohlich2004another,merkli2020quantum}. These approaches establish equilibration through spectral estimates, resonance expansions, and long-time bounds under suitable assumptions, but do not reveal how finite-coupling stationary-state correction is encoded within the reduced dynamics.

The first finite-coupling corrections to the stationary state appear at second order in the system--bath coupling. At this point, the question changes from \emph{How close does the system approach the equilibrium state?} to \emph{Why does the reduced dynamics know how to construct the Gibbs correction?} This mean-force Gibbs correction~\cite{cresser2021weak} is thus not merely a higher-precision characterization of equilibrium, but also a probe of how finite-coupling stationary-state information is encoded within the reduced dynamics. Accordingly, second-order return to equilibrium (R2E$_2$) is a nontrivial finite-coupling stationary-state problem with an independently known stationary-state correction.

Progress toward understanding the finite-coupling stationary-state corrections has been limited. The TCL2 generator correctly reproduces the equilibrium second-order coherence corrections~\cite{divincenzo2005rigorous,Fleming,Thingna,Guarnieri2018,tupkary2021fundamental}, but its stationary populations remain accurate only to zeroth order~\cite{Fleming,Thingna,tupkary2021fundamental}. Crowder \textit{et al.} demonstrated numerically that the TCL4 stationary population corrections reproduce the equilibrium corrections at zero temperature for arbitrary finite-dimensional systems and sub-Ohmic through super-Ohmic baths~\cite{Crowder}. Kumar, Athulia, and Ghosh provided the first analytic proof of stationary state corrections for a qubit coupled to a bath with odd spectral density~\cite{kumar2024equivalence}, demonstrating extensive cancellations among the contributing terms, while Lampert \textit{et al.} recovered the equilibrium correction through simpler partial resummation for arbitrary spectral densities and restricted qubit couplings~\cite{Lampert2025}.

A complementary observation was made earlier by Thingna, Wang, and H\"anggi, who showed that the mean-force Gibbs population correction can be obtained by analytic continuation of stationary second-order coherence corrections of the Redfield (TCL2) master equation~\cite{Thingna}. The desired correction therefore appears to be encoded in the in TCL2 coherence data.  Becker \textit{et al.} connected such continuations to canonically consistent closures  incorporating the anticipated
finite-coupling equilibrium corrections directly into the
master equation~\cite{becker2022canonically},
although their origin, uniqueness, and range of validity
remain unclear. Tupkary \textit{et al.} emphasized that second-order
perturbation theory alone is insufficient to
capture finite-coupling population dynamics,
necessitating explicit fourth-order population
generators to determine these
corrections~\cite{tupkary2021fundamental}.

Taken together, these results point toward an underlying organization of finite-coupling stationary states in terms of simplified dynamical building blocks hidden within the asymptotically redundant TCL4 generator.

The challenge is that this organization is
already encoded in the van-Kampen cumulant expansion of
time-convolutionless (TCL)
dynamics~\cite{VANKAMPEN1,
ChaturvediShibata1979,
BreuerHeinz-Peter1961-2007TToO}. The cumulants are notoriously complicated
objects, so their asymptotic organization has
remained largely implicit. Attempts to organize this complexity through recursive relations, algebraic approaches, and Penrose-inverse methods have been made~\cite{gasbarri2018recursive,trushechkin2021derivation,karasev2023timeconvolutionless,Blumenfeld2025PseudoinverseTCL}. Nevertheless, a transparent description of higher-order TCL dynamics remains elusive. This has favored simplified effective descriptions,
including
Gorini--Kossakowski--Sudarshan--Lindblad
(GKSL)~\cite{lindblad1976,Gorini1976}
and Redfield-type~\cite{Redfield1957,Hartmann00}
approaches, where the rules governing the dynamics are
more transparent, albeit at the cost of reduced or
uncontrolled accuracy. Examples include
coarse-grained master
equations~\cite{Majenz,Schaller},
universal~\cite{Nathan,Davidovic2020,fernandez2024recovering}
and unified~\cite{Vogt,Tscherbul,Jeske,Trushechkin}
Lindblad equations, and canonically consistent
closures~\cite{tupkary2023searching,becker2022canonically}.

The existence of KMS-induced cancellations helps explain
the TCL4 asymptotic redundancy. At fourth
order, only the contraction of the TCL4 generator with
the zeroth-order stationary state enters the
correction. Consequently, generator components that
annihilate the zeroth-order stationary state are
asymptotically invisible. Distinct TCL4
representations may therefore produce the same
asymptotic correction. We show that these
representations are connected by stationary-state-preserving
transformations that progressively reduce the complexity
of the faithful TCL4 generator while preserving the
stationary state.

The first and largest step in this reduction is an
asymptotic selection rule, which removes the most
complicated nonseparable TCL4 contributions. The
remaining latent-resonance generator retains the
finite-coupling stationary information while admitting
a closed analytic form.

The latent-resonance generator further admits an auxiliary
representation, which yields a
triangle identity relating three generators through
stationary-state-preserving transformations. One vertex of this triangle is the virtual coherence pathway, determined solely in terms of the TCL2-generator, showing that the explicit TCL4 population generator is ultimately unnecessary for determining the second-order stationary-state corrections. 

\subsection{Main Result and Paper Outline}

Consider an open quantum system coupled weakly
(\(\lambda^2\ll1\)) to a single thermal bath at inverse
temperature \(\beta\). 
Let \(H_S\) denote the system
Hamiltonian with nondegenrate energy spectrum
\(\{E_n\},n=1,\ldots,N
\).
Let
\[
L^{(0)}+\lambda^2L^{(2)}
\]
denote the Redfield generator, where
\(L^{(0)}=-i[H_S,\cdot]\).

For systems whose Bohr frequencies exceed all
relaxation rates, the state remains accurate to
\(O(\lambda^2)\) under the simple canonically
consistent master equation
\begin{equation}
\frac{d\rho}{dt}
=
\bigl(L^{(0)}+\lambda^2L^{(2)}\bigr)\rho
+
\lambda^4\,\mathrm{diag}(\bar V p^{(0)}),
\label{Eq:MainResult}
\end{equation}
where the last term denotes a diagonal matrix with the vector $\bar Vp^{(0)}$ on the diagonal.
The corrected equation is
Hermiticity preserving and trace preserving.

Here \(p^{(0)}\) denotes the zeroth-order asymptotic
population vector of the Redfield equation, and
\(\bar Vp^{(0)}\) denotes the full virtual coherence pathway contraction
\begin{equation}
(\bar V p^{(0)})_n
=
-\sum_{a,b=1}^{N}
L^{(2)}_{nn,ab}
\,c^{(2)}_{ab}.
\label{Eq:mainSum}
\end{equation}
The second-order stationary coherence corrections are
\begin{equation}
c^{(2)}_{ab}
=
\frac{-i}{E_a-E_b}
\sum_{m=1}^{N}
L^{(2)}_{ab,mm}\,
p_m^{(0)}.
\end{equation}
The diagonal contributions are included in
Eq.~\eqref{Eq:mainSum} through an auxiliary (extended)
Hilbert-space extension of Sec.~\ref{Sec:AuxCover} and correspond to the
analytic-continuation procedure from coherences to
populations introduced by Thingna
\emph{et al.}~\cite{Thingna},
\begin{equation}
c^{(2,{\rm aux})}_{nn}
=
\lim_{E_k\to E_n}
c^{(2,{\rm aux})}_{nk}.
\end{equation}
The auxiliary Hilbert-space construction provides a
systematic realization of this continuation procedure.
It enables populations to couple to auxiliary
coherences even when population--coherence couplings
are forbidden in the physical system. 

Upon pullback to the physical system (which amounts to the
appropriate rescaling of the auxiliary variables), the resulting
diagonal contributions satisfy
\[
c^{(2)}_{nn}=p_n^{(2)},
\]
where \(p_n^{(2)}\) is the second-order stationary-state
population correction.

The remainder of the paper derives this result starting
from the TCL4 population generator through a sequence of
stationary-state-preserving transformations. The equilibrium correction serves primarily as a guide
for identifying the selection rule, which in turn
explains why TCL4 reproduces it as the stationary state.  The appendices
provide the analytical and numerical machinery
underlying these results.

\section{Open Quantum System Model}

We consider a finite-dimensional system coupled linearly
to a single thermal bath,
\[
H=H_S+H_B+\lambda H_I,
\qquad
H_I=A\otimes B ,
\]
where
\[
H_S=\sum_{n=1}^N E_n |n\rangle\langle n|.
\]

We consider the compact space \(\mathcal H_M\) of bounded
system Hamiltonians satisfying
\[
\|H_S\|\le M .
\]
Throughout this work we restrict attention to
perturbative orders for which the asymptotic TCL
generator exists. The
asymptotic generator depends continuously
on \(H_S\) throughout \(\mathcal H_M\).

The subset
\[
\mathcal H_{\rm nd}
=
\{H_S\in\mathcal H_M:\ E_n\neq E_m
\text{ for } n\neq m\}
\subset \mathcal H_M
\]
of nondegenerate Hamiltonians is open and dense in
\(\mathcal H_M\), while degenerate Hamiltonians lie on
its boundary.

The bath consists of harmonic
modes,
\[
H_B=\sum_k \omega_k b_k^\dagger b_k,
\]
and couples linearly to the system through
\[
H_I=A\otimes B,
\qquad
B=\sum_k g_k(b_k+b_k^\dagger),
\]
where \(A\) is Hermitian,
\(g_k\propto\lambda\ll1\),
and \(b_k\) are bosonic annihilation operators.

The reduced dynamics are governed by the thermal bath
correlation function
\[
C(t)=\langle B(t)B(0)\rangle,
\]
evaluated on the bath Gibbs state at inverse
temperature \(\beta\). 
It admits the Fourier representation
\[
C(t)
=
\frac{1}{\pi}
\int_{-\infty}^{\infty}
d\omega\,
J_\omega e^{-i\omega t}.
\]
The thermal spectral density satisfies the KMS relation
\[
J_{-\omega}=e^{-\beta\omega}J_\omega .
\]

A central quantity is the half-sided transform
\[
\Gamma_\omega
=
\int_0^\infty
dt\, C(t)e^{i\omega t}
=
J_\omega+iS_\omega,
\]
together with its finite-time counterpart
\[
\Gamma_\omega(t)
=
\int_0^t
d\tau\, C(\tau)e^{i\omega\tau}.
\]
The long-time limit
\[
\Gamma_\omega(t)\rightarrow\Gamma_\omega
\]
generates the asymptotic TCL coefficients, while the
finite-time transform encodes the non-Markovian
 dynamics.
\(\Gamma_\omega\) is analytic in the upper-half
frequency plane.

\subsection{Time-Convolutionless Generators and Mean-Force Gibbs Corrections}

The system density matrix admits the time-local equation
\begin{equation}
\frac{d \rho_S(t)}{dt}=L(t)\rho_S(t),
\end{equation}
where \(L(t)\) is the TCL generator, which is represented as
\[
\dot{\rho}_{nm}
=
\sum_{ij}L_{nm,ij}(t)\rho_{ij},
\]
in the isolated system energy eigenbasis,

The dimensionless weak-coupling parameter \(\lambda^2\)
is factored outside the bath correlation functions and
all associated transforms. Accordingly, the TCL
generator admits the expansion
\begin{equation}
L(t)=L^{(0)}
+\lambda^2L^{(2)}(t)
+\lambda^4L^{(4)}(t)
+O(\lambda^6).
\label{Eq:TCL}
\end{equation}

In the long-time limit we write
\[
L_\infty
=
L^{(0)}
+\lambda^2L^{(2)}
+\lambda^4L^{(4)}
+O(\lambda^6),
\]
\[
L^{(2k)}=\lim_{t\to\infty}L^{(2k)}(t),
\qquad
k=1,2.
\]
The free generator matrix elements are
\[
L^{(0)}_{nm,ij}
=
-i(E_n-E_m)\delta_{ni}\delta_{mj}.
\]
The second order (TCL2) matrix elements are
\begin{align}
L^{(2)}_{nm,ij}
&=
A_{ni}A_{jm}
\left(
\Gamma_{in}+\Gamma_{jm}^{*}
\right)
\nonumber\\
&\quad
-
\delta_{mj}
\sum_k
A_{nk}A_{ki}\Gamma_{ik}
-
\delta_{ni}
\sum_k
A_{jk}A_{km}\Gamma_{jk}^{\ast}.
\label{Eq:TCL2General}
\end{align}
This gives the standard Redfield equation.

Only the TCL4 population block is required to achieve
second-order corrections of the stationary
state~\cite{Fleming}. We begin with Crowder's form of the TCL4
generator~\cite{Crowder}. Restricting to the
asymptotic population block yields a sum
whose terms couple three Bohr
frequencies,
\begin{widetext}
\begin{equation}
\begin{aligned}
L^{(4)}_{nn,ii}
&=
\sum_{a,b,c}
A_{ia}A_{ab}A_{bc}A_{ci}\,\delta_{ni}
\Big[
F_{cb,ci,ac} 
-
R_{cb,ab,bi}
-
F_{ba,ci,ac}
+
R_{ic,ab,bi}
\Big],
\\[4pt]
&\quad
+
\sum_{a,b}
A_{na}A_{ab}A_{bi}A_{in}
\Big[
-F_{ba,bi,nb}
+
R_{ba,na,ai}
+
F_{an,bi,nb}
-
R_{ib,na,ai}
\Big]
\\
&\quad
+
\sum_{a,b}
A_{na}A_{ab}A_{bi}A_{in}
\Big[
C_{ba,in,ai}
+
R_{ba,in,ai}
-
C_{ib,in,ai}
-
R_{ib,in,ai}
\Big]
\\
&\quad
+
\sum_{a,b}
A_{na}A_{ai}A_{ib}A_{bn}
\Big[
-C_{an,ib,ni}
-
R_{an,ib,ni}
+
C_{ia,ib,ni}
+
R_{ia,ib,ni}
\Big]
+\text{c.c.}
\end{aligned}
\label{Eq:TCL4Population}
\end{equation}
\end{widetext}
with \(\text{c.c.}\) indicating the complex conjugate.
The three-frequency kernels are
\begin{align}
F_{\omega_1\omega_2\omega_3}
&=
i\,\Gamma^{T}_{\omega_2}
\frac{
\Gamma_{-\omega_2-\omega_3}
-
\Gamma_{\omega_1}
}{
\omega_1+\omega_2+\omega_3
},\label{Eq:TCLF}
\\[8pt]
C_{\omega_1\omega_2\omega_3}
&=
i\,\Gamma^{\star}_{\omega_2}
\frac{
\Gamma_{-\omega_2-\omega_3}
-
\Gamma_{\omega_1}
}{
\omega_1+\omega_2+\omega_3
},\label{Eq:TCLC}
\\[8pt]
R_{\omega_1\omega_2\omega_3}
&=
\tilde{R}_{\omega_1\omega_2\omega_3}
+
i\,\Gamma_{\omega_2}
\frac{
\Gamma_{-\omega_2-\omega_3}
-
\Gamma_{\omega_1}
}{
\omega_1+\omega_2+\omega_3
},\label{Eq:TCLR}
\end{align}
where \(\tilde R\) admits both the time-domain and
frequency-domain representations
\begin{subequations}\label{Eq:RtildeBoth}
\begin{align}
\tilde{R}_{\omega_1\omega_2\omega_3}
&=
-\int_0^\infty dt\
\Delta\Gamma_{\omega_1}(t)
\Delta\Gamma_{\omega_2}(t)
e^{-i\Omega t}
\label{Eq:RtildeTime}
\\
&=
\frac{1}{\pi}
\int_{-\infty}^{\infty}
d\omega\,
J_\omega
\frac{
\Gamma_{\omega_1}
-
\Gamma_{-\omega_3-\omega}
}{
(\omega-i0^+-\omega_2)
(\omega_1+\omega_3+\omega)
}.
\label{Eq:RtildeFreq}
\end{align}
\end{subequations}
Here 
\begin{equation}
\Omega=\omega_1+\omega_2+\omega_3,\label{Eq:Omega}
\end{equation}
and
\begin{equation}
\Delta\Gamma_\omega(t)=\Gamma_\omega-\Gamma_\omega(t).
\end{equation}

Equation~\eqref{Eq:TCL4Population} is a faithful representation of the TCL4
population dynamics. The kernel functions couple
three Bohr frequencies through nonseparable frequency denominators and frequency integrals, producing highly nontrivial frequency combinatorics. While the asymptotic constraints derived
below are fully encoded in this expression, their
organization is not manifest.

\subsubsection*{Second-Order Return to Equilibrium and the Simplicity of Mean-Force Gibbs Corrections}

Return to equilibrium may be formulated at several
levels~\cite{trushechkin2022open}.
In this paper, the exact R2E condition is
\begin{equation}
\lim_{t\to\infty}\rho_S(t)
=
\frac{\mathrm{Tr}_B\!\left(e^{-\beta H}\right)}
{\mathrm{Tr}\!\left(e^{-\beta H}\right)}.
\label{Eq:ExactR2E}
\end{equation}

In the weak-coupling regime, one may consider
the perturbative second-order return-to-equilibrium
condition
\begin{align}
p
&=
p^{(0)}
+
\lambda^2 p^{(2)}
+
O(\lambda^4),
\\
c
&=
\lambda^2 c^{(2)}
+
O(\lambda^4),
\label{Eq:R2E2}
\end{align}
where \(p\) and \(c\) denote the populations and
coherences of the reduced state.

The R2E$_2$ condition requires that the
corrections \(p^{(2)}\) and \(c^{(2)}\) obtained from
the stationary state
coincide with the corresponding corrections obtained
from the  equilibrium
state expansion.

The equilibrium coherences are reproduced
directly by the TCL2 population-to-coherence matrix
elements. The nontrivial part of the
R2E$_2$ problem is in finding the
stationary state population correction~\cite{Fleming,tupkary2021fundamental}.

To distinguish between the stationary and equilibrium
populations, we reserve the symbols \(p\) and \(q\),
respectively, throughout the remainder of the paper.
The nondegenerate perturbation theory of the mean-force
Gibbs state~\cite{cresser2021weak} yields the
second-order equilibrium populations as a 
sum of terms depending on a single Bohr
frequency,\footnote{
The principal density \(A_\beta(\omega)\) in
Ref.~\cite{cresser2021weak} is related to
\(S(\omega)\) in the present notation through
\(A_\beta(\omega)=-S(\omega)\).
The normalization contribution proportional to
\(p_n^{(0)}\) has been omitted since it plays no role
in stationary condition.
}
\begin{equation}
\label{Eq:mfgscor}
\begin{split}
q_n^{(2)}
&=
\sum_{k=1}^N\delta q_{nk}^{(2)},\\
\delta q_{nk}^{(2)}
&=
|A_{kn}|^2\big(
q_n^{(0)}S'_{nk}
-q_k^{(0)}S'_{kn}
-\beta q_n^{(0)}S_{nk}
\big).
\end{split}
\end{equation}

This expression makes the central question of the paper explicit: why does the intricate TCL4 population generator reproduce the comparatively simple mean-force Gibbs correction?

\subsection{Germs and Decompositions}

A germ is a minimal analytically independent contribution to a generator. It is distinguished from the more general notion of a term by its linear independence and is therefore analogous to a basis vector. Each germ contains a particular product of coupling matrix elements, spectral functions, principal-value functions, and their derivatives evaluated at specific Bohr frequencies. Distinct germs are identified by the independent analytic entities they contain.

Schematically, a germ has the product form
\[
T_\alpha
\sim
A_{\alpha_1}\cdots A_{\alpha_m}
\,J(\omega_{\alpha_1})
\,S(\omega_{\alpha_2})
\,J'(\omega_{\alpha_3})
\,S'(\omega_{\alpha_4}),
\]
where the frequencies and matrix elements depend on the label
\(\alpha\). For example, Eq.~\eqref{Eq:mfgscor} contains three
germs. Two belong to the same derivative germ family,
while the third belongs to a distinct principal-value
family.

Because the matrix elements, spectral functions, and frequencies may
be varied independently, distinct germs are analytically independent.
Consequently, identities expressed in terms of independent germ products may be decomposed into identities for the corresponding germ families.

 Consider an identity of the form
\begin{equation}
\label{Eq:Germstotal}
\sum_{\alpha\in\mathcal F_1}T_\alpha
+
\sum_{\beta\in\mathcal F_2}T_\beta
=0,
\end{equation}
where \(\mathcal F_1\cap\mathcal F_2=\varnothing\). Linear independence then
implies
\begin{equation}
\sum_{\alpha\in\mathcal F_1}T_\alpha=0,
\qquad
\sum_{\beta\in\mathcal F_2}T_\beta=0.
\label{Eq:Germstratified}
\end{equation}

\section{Perturbative Stationary State and Stationary-State-Preserving Generators\label{Sec:PST}}

Let the superoperator \(P\) project onto populations,
\[
P\rho
=
\sum_n \rho_{nn}|n\rangle\langle n|,
\]
and let \(Q=1-P\) project onto coherences.
Any superoperator \(X\) then decomposes into blocks
\[
X_{AB}=AXB,
\qquad
A,B\in\{P,Q\},
\]
mapping between the projected subspaces.

The block \(L^{(0)}_{QQ}\) is diagonal and invertible on
\(\mathcal H_{\rm nd}\),
\begin{equation}
\big[\big(L^{(0)}_{QQ}\big)^{-1}X\big]_{mn}
=
\frac{iX_{mn}}{E_m-E_n}.
\label{Eq:poles}
\end{equation}
As Bohr frequencies approach zero, the inverse develops
simple poles. In contrast, the TCL generators
\(L^{(2)}\) and \(L^{(4)}\) remain analytic and admit
continuous continuation to the compact space
\(\mathcal H_M\).

To determine the asymptotic states within
\(\mathcal H_{\rm nd}\), we deploy the nondegenerate
perturbation theory of stationary states~\cite{Fleming,tupkary2021fundamental}. The
nondegenerate expansion is valid when the energy gaps
are large compared with the relaxation
rates generated by the system--bath interaction.

 The stationary  populations and coherences are expanded as,
\begin{align}
p_\infty
&=
p^{(0)}
+\lambda^2p^{(2)}
+O(\lambda^4),
\qquad
p^{(0)}_n
=
\frac{e^{-\beta E_n}}{Z_S},
\\
c_\infty
&=
c^{(0)}
+\lambda^2c^{(2)}
+O(\lambda^4),
\qquad
c^{(0)}_{nm}=0,
\quad n\neq m .
\end{align}
Here we use the standard result that the zeroth-order
stationary state coincides with the Gibbs state for a
system coupled to a single thermal bath~\cite{davies1974,Fleming,tupkary2021fundamental}.

Expanding the stationary equation \(L\rho_\infty=0\) to second order
in the populations and eliminating the coherence block gives the
cyclic Gibbs condition
\begin{equation}
L^{(2)}_{PP}p^{(0)}=0
\label{Eq:AST0}
\end{equation}
and the correction equation
\begin{equation}
L^{(2)}_{PP}p^{(2)}
=\Big[
-
L^{(4)}_{PP}
+
L^{(2)}_{PQ}
\big(L^{(0)}_{QQ}\big)^{-1}
L^{(2)}_{QP}\Big]p^{(0)}.
\label{Eq:ASTp}
\end{equation}
Thus the second-order population correction is controlled
by two effective fourth-order contributions: the TCL4
population generator and the virtual coherence pathway,
\begin{equation}
V\equiv
L^{(2)}_{PQ}
\big(L^{(0)}_{QQ}\big)^{-1}
L^{(2)}_{QP},
\qquad
H_S\in\mathcal H_{\rm nd},
\label{Eq:VCPdef}
\end{equation}
which is defined only on the reduced nondegenerate
domain \(\mathcal H_{\rm nd}\). 
The matrix elements of $V$ are
\begin{equation}
 V_{nn,ii}
=
i\sum_{\substack{a,b=1\\\omega_{ab}\neq 0}}^N
\frac{L^{(2)}_{nn,ab}L^{(2)}_{ab,ii}}{\omega_{ab}}
\label{Eq:VCPdefME}
\end{equation}
$V$ inherits the poles in
Eq.~\eqref{Eq:poles} and becomes singular near the
boundary of \(\mathcal H_{\rm nd}\), signaling the breakdown of
the nondegenerate perturbative theory.

Similarly, eliminating the population block gives
\begin{equation}
c^{(2)}
=
-\big(L^{(0)}_{QQ}\big)^{-1}
L^{(2)}_{QP}p^{(0)} ,
\label{Eq:c2}
\end{equation}
for the second-order coherence correction. Combining Eqs.~\eqref{Eq:VCPdefME} and~\eqref{Eq:c2} we find
\begin{equation}
Vp^{(0)}=
-L^{(2)}_{PQ}
c^{(2)}.
\label{Eq:VCPdef2}
\end{equation}

The stationary-state population depend on the fourth-order population generator only through the Gibbs contraction appearing in Eq.~\eqref{Eq:ASTp}. Consequently, different fourth-order generators may yield the same second-order stationary-state correction. For example, the transformation \[ \lambda^4 L^{(4)} \rightarrow \lambda^4\big [L^{(4)}+cL^{(2)}\big] \] leaves the correction \(p^{(2)}\) invariant because of Eq.~\eqref{Eq:AST0}. More generally, two fourth-order population generators are said to be \emph{stationary-state-preserving}, if they produce the same Gibbs contraction, namely
\begin{equation}
\label{Eq:EqEquiv}
\begin{aligned}
L^{(4)\prime}
&\xleftrightarrow{\;\mathcal T\;}
L^{(4)},
\\
L^{(4)\prime}-L^{(4)}
&=\mathcal T,
\qquad
\mathcal T p^{(0)}=0.
\end{aligned}
\end{equation}
The operator \(\mathcal T\) will be referred to as a
\emph{stationary-state-preserving transformer}.

The main thesis of the present work is to exploit this
equivalence to simplify the TCL4 generator and uncover
how its stationary-state corrections are organized. In the framework developed below, we will encounter
stationary-state-preserving transformations repeatedly below,
in the TCL4 selection rule, the elimination of principal integrals and in
the auxiliary extensions.

\subsection{Resonant-Nonresonant Decomposition}
Next, we decompose the TCL4 population generator~\eqref{Eq:TCL4Population} into
nonresonant and latently resonant components,
\[
L^{(4)}_{PP}
=
L_{\rm rg}
+
N_{\rm rg}.
\]
This decomposition is purely algebraic and remains well defined on
the compact space \(\mathcal H_{\rm M}\).

The latent resonance generator \(L_{\rm rg}\) consists of
all TCL4 contributions satisfying the latent resonance
condition \(\Omega=0\),
\begin{equation}
L_{\rm rg}
\equiv
\sum_{\Omega=0}
L^{(4)}_{\Omega},
\label{Eq:resonantselection}
\end{equation}
whereas the nonresonant remainder can be written as
\begin{equation}
N_{\rm rg}
\equiv
\sum_{\Omega\neq 0}
L^{(4)}_{\Omega}.
\label{Eq:nresonantselection}
\end{equation}
The decomposition is defined algebraically through the
kernel representation of
Eq.~\eqref{Eq:TCL4Population} by the exact condition
\(\Omega=0\). 

Unlike ordinary energy degeneracies,
these resonances occur between Bohr frequencies and
therefore persist even in nondegenerate systems. We
therefore refer to them as latent.

Because of the constraint \(\Omega=0\),
\(L_{\rm rg}\) contains only two-frequency
interactions. As we show below, these interactions
become separable and admit an expansion in terms of
germs, which include the derivatives of spectral
functions appearing in the mean-force Gibbs
corrections~\eqref{Eq:mfgscor}. By contrast,
\(N_{\rm rg}\) retains the nonseparable
three-frequency interactions of the TCL4 generator and
does not admit such a germ expansion. The selection rule
introduced next isolates the latent-resonance generator
as the sole contributor to the stationary-state
equation~\eqref{Eq:ASTp1}.

\subsection{TCL4 Selection Rule}

The stationary state condition can now be recast as
\begin{equation}
L^{(2)}_{PP}p^{(2)}
=\big(
-
L_{rg}-N_{rg}
+
V\big) p^{(0)}.
\label{Eq:ASTp1}
\end{equation}

Agreement between the stationary state of the TCL4
generator and the mean-force Gibbs correction requires
an asymptotic selection rule relating the independent
analytic components entering the R2E$_2$
condition~\eqref{Eq:ASTp1}. The mean-force Gibbs
correction admits an expansion in derivative spectral
germs, whereas the interior representations of
\(N_{\rm rg}\) and \(V\) contain no derivative
contributions.

This requires the selection rule
\begin{equation}
(N_{\rm rg}-V)p^{(0)}=0,
\label{Eq:SelectionRule}
\end{equation}
which identifies \(N_{\rm rg}-V\) as the
stationary-state-preserving transformer introduced in
Sec.~\ref{Sec:PST}.

The selection rule is proven analytically for a single
qubit and confirmed numerically for larger \(N\) in
Appendix~B.
Although a general analytic proof for arbitrary \(N\) is
not presently available because of the extreme complexity
of \(N_{\rm rg}\), the extensive numerical evidence
presented in Appendix~B strongly supports its validity,
and we therefore assume that the selection rule holds for
arbitrary finite-dimensional systems.

The selection rule is a necessary condition for R2E$_2$, following from the linear independence of the analytic components entering the TCL4 stationary-state condition. While R2E$_2$ 
 motivates the identification of Eq.~\eqref{Eq:SelectionRule}, the remainder of this work
treats it as an intrinsic constraint of the TCL4
generator in its own right.

An immediate corollary is that the stationary Eq.~\eqref{Eq:ASTp1} can be simplified to
\begin{equation}
L^{(2)}_{PP}p^{(2)}
=
-
L_{\rm rg} p^{(0)}.
\label{Eq:ReducedAST}
\end{equation}
Thus, the selection rule removes the 
virtual coherence pathway from the 
stationary-state equation. The reduced equation depends only on 
the components of \(L^{(2)}\) and \(L^{(4)}\), which are bounded on \(\mathcal H_M\), and is now independent of
the virtual coherence pathway \(V\), which develops
poles as energy levels approach degeneracy.

Importantly, the selection rule should not be confused
with the validity of nondegenerate perturbation theory.
Near degeneracies, neither \(p^{(2)}\) nor the
mean-force Gibbs correction can, in general, be obtained
from the nondegenerate expansion. The selection rule is
not a statement about a particular perturbative
approximation, but an algebraic identity between
generators. It remains valid near degeneracies, where
the virtual coherence pathway diverges (though not its
stationary state contraction). The validity of the perturbative
expansion and the validity of the selection rule are
distinct issues.

We next turn to a near-degenerate spectrum, where the
algebraic identity~\eqref{Eq:SelectionRule} can be
investigated through its decomposition into boundary and
interior contributions. The Bohr
frequencies separate into two subsets: those smaller than
an arbitrary scale \(\epsilon\), and those that approach
nonzero values as \(\epsilon\to0\). We define the
boundary operator \(\partial_\epsilon\) by its action on
\(N_{\rm rg}\) and \(V\):
\begin{align}
\partial_\epsilon N_{\rm rg}
&=
\sum_{0<|\Omega|<\epsilon}
L^{(4)}_{\Omega},
\label{Eq:nresonantselectionBC}
\\
(\partial_\epsilon V)_{nn,ii}
&=
i\sum_{\substack{a,b=1\\0<|\omega_{ab}|<\epsilon}}^N
\frac{
L^{(2)}_{nn,ab}L^{(2)}_{ab,ii}
}{\omega_{ab}}.
\label{Eq:VCPdefME1}
\end{align}
In the limit \(\epsilon\to0\), the selection rule
stratifies into two independent identities,
\begin{align}
\partial N_{\rm rg}p^{(0)}
&=
\partial Vp^{(0)},
\label{Eq:LocalBoundaryLaw}
\\
(1-\partial)N_{\rm rg}p^{(0)}
&=
(1-\partial)Vp^{(0)}.
\label{Eq:LocalBoundaryComplement}
\end{align}
The existence of the limit
\(\partial Vp^{(0)}\) follows from the existence of the
stationary coherence analytic continuation
~\cite{Thingna}. The separation into two independent identities follows
from the linear independence of the derivative germ
families associated with the boundary and the
nonderivative interior contributions. An illustrative example is given in Appendix~\ref{Ap:stratificationexample}.

\section{Standard Form of the Latent-Resonance Generator}

The next step simplifies the latent-resonance generator in Eq.~\eqref{Eq:resonantselection} through a further stationary-state-preserving transformation by bringing it into a separable form.

Appendix~\ref{Ap:PIelimination} shows that the principal-value contributions generated by 
\(\tilde R\) through Eq.~\eqref{Eq:RtildeFreq} define an stationary-state-preserving transformation. Eliminating these contributions brings the latent-resonance generator into the standard form,
\begin{equation}
L_{nn,ii}^{\rm std}
=
L_{nn,ii}^{\rm qu}
+
L_{nn,ii}^{\rm int}.
\end{equation}
The off-diagonal matrix elements, \(n\neq i\), are

\begin{widetext}
\begin{subequations}\label{Eq:StandardForm}
\begin{align}
L_{nn,ii}^{\rm qu}
&=
2\vert A_{ni}\vert^4
\big[
(J_{in}-J_{ni})S'_{in}
+
(S_{in}-S_{ni})J'_{in}
+
2J_{in}S'_{ni}
\big]
+
2\vert A_{ni}\vert^2(A_{ii}^2-A_{nn}^2)S_0,
\label{Eq:SFembed}
\\
L_{nn,ii}^{\rm int}
&=
\sum_{k\neq n, i}
\bigg[
\vert A_{nk}\vert^2
\big(
2\vert A_{ni}\vert^2J_{in}S'_{nk}
-
\vert A_{ik}\vert^2J_{ik}S'_{kn}
+
\vert A_{ik}\vert^2J_{kn}S'_{ik}
\big)
\nonumber\\
&\qquad
-
i\vert A_{ni}\vert^2\Gamma'_{in}
\big(
\vert A_{ik}\vert^2\Gamma_{ik}
-
\vert A_{nk}\vert^2\Gamma_{nk}
\big)
\bigg]
+\mathrm{c.c.}
\label{Eq:StandardFormFinal}
\end{align}
\end{subequations}
\end{widetext}
The diagonal elements are obtained from trace
preservation, 
\[
L_{nn,nn}^{\rm std}
=
-\sum_{i\neq n}L_{ii,nn}^{\rm std}.
\]

The term \(L^{\rm qu}\) will be referred to as the
\emph{qubit term}. It is local to the transition
frequency \(\omega_{ni}\). The \emph{interaction term}
\(L^{\rm int}\) contains pairwise interactions between
distinct Bohr frequencies and does not annihilate the
zeroth-order stationary state, implying that the interactions participate
in the stationary state corrections. Moreover, the
interaction terms factorize between pairs of 
frequencies.

Thus, the stationary state is governed by the interplay between local
qubit contributions associated with individual Bohr
frequencies and bath-mediated interactions among distinct
Bohr frequencies.

\subsection{Equivalence of the Equilibrium and Stationary States}
We now show that the solution of
Eq.~\eqref{Eq:ReducedAST} is given by the second-order
equilibrium correction,
\begin{equation}
p^{(2)} = q^{(2)},
\label{Eq:R2E2m}
\end{equation}
where \(q^{(2)}\) is defined by
Eq.~\eqref{Eq:mfgscor}. The detailed proof proceeds by induction
and is deferred to Appendix~\ref{App:QubitR2E}.

The induction
begins with the case \(N=2\), where \(L_{\rm rg}\)
contains only the qubit term. We find
\begin{equation}
L^{(2)}_{PP}q^{(2)}
=
-
L_{\rm qu}p^{(0)}.
\label{Eq:QubitR2E}
\end{equation}
Together with the proof of the TCL4 selection rule for the qubit, in
Appendix~\ref{App:SelectionRule}, this establishes the
equivalence between the TCL4 stationary state and the
mean-force Gibbs correction for the qubit. The result
previously obtained in
Ref.~\cite{kumar2024equivalence} for baths with odd
spectral densities is thus recovered here for arbitrary bath
spectral densities.

Next, we assume that Eq.~\eqref{Eq:R2E2m}
holds for an \(N\)-level system and show that the
increment to an \(N+1\)-level system also satisfies the
same identity. 
The induction increment is supported on the set of
all pairwise interactions involving the added level.
We refer to this set as the extension shell,
\begin{equation}
\mathcal S_{N+1}
=
\{(N+1,i)\;|\; i=1,\ldots,N\}.
\label{Eq:ExtensionShell}
\end{equation}
The induction step separates into the
qubit term \(L_{\rm qu}\) and the interaction
term \(L_{\rm int}\). 
The qubit term cancels
locally on the extension shell. Separability 
ensures that the interaction term contraction produces precisely the new
pairwise mean-force Gibbs contributions involving the
added level.

Thus, the latent-resonance
generator reproduces exactly the induction increment of
the mean-force Gibbs correction. By induction,
Eq.~\eqref{Eq:R2E2m} holds for arbitrary \(N\),
establishing the sufficiency of the TCL4 selection rule for R2E$_2$. Together with the necessity established above, this proves the equivalence between the TCL4 selection rule and R2E$_2$.

\section{Auxiliary Cover and Boundary Representation\label{Sec:AuxCover}}

Although substantially simpler than the faithful TCL4 population generator, the latent-resonance generator remains a nontrivial object. The auxiliary representation introduced below provides a further reduction in complexity and a more transparent characterization of its stationary-state content.

Latent resonances are resonances between Bohr
frequencies that occur even in nondegenerate systems.
To relate the latent-resonance generator to the more
familiar energy degeneracies, we begin with the
observation that the induction step used above to
establish Eq.~\eqref{Eq:R2E2m} for arbitrary \(N\) may
also be viewed as the addition of an auxiliary level. In the induction proof, the added
level changes the system by introducing new
Bohr frequencies and new interactions. However, it
is also possible to introduce auxiliary levels without
changing the system. This is
achieved by creating near-degenerate replicas of existing levels and
their associated transition matrix elements.

By introducing such replica levels, latent resonances are unfolded into explicit boundary representatives obtained by analytic continuation from the interior of  $\mathcal H_{\rm nr}^{\rm aux}$. The resulting collection of boundary germs defines an auxiliary cover of the latent-resonance germ family: each latent-resonance germ admits an analytic continuation representative, and each boundary germ projects back onto a latent-resonance germ.

The complete auxiliary cover is a boundary realization of the equivalence class of $L_{rg}$
under stationary-state-preserving transformations.

\subsection{Complete Auxiliary Cover}

This cover is a replicated \(2N\)-dimensional auxiliary
space,
\[
\mathcal H_{\rm aux}
=
\{E_1,\ldots,E_N,E_{N+1},\ldots,E_{2N}\},
\]
where each auxiliary level \(E_{N+n}\) is nearly degenerate with \(E_n\). 
Unless otherwise stated, the auxiliary covers are understood at finite but arbitrarily small detuning, with the degenerate limit taken only at the end of the calculation.

The coupling operator
is enlarged to
\begin{equation}
A^{(\rm a)}
=
\frac{1}{\sqrt[4]{2}}
\begin{pmatrix}
A & A\\
A & A
\end{pmatrix},
\label{Eq:Aaux}
\end{equation}
which fixes the normalization of the auxiliary generators.

At strict degeneracy, the full auxiliary Hamiltonian
\(H^{\rm aux}\) possesses an exchange symmetry,
\[
[H^{\rm aux},R\otimes I_B]=0,
\qquad
R=
\begin{pmatrix}
0 & \mathbf 1\\
\mathbf 1 & 0
\end{pmatrix},
\]
where \(\mathbf 1\) denotes the \(N\times N\) identity
matrix and \(I_B\) is the identity on the bath Hilbert
space. Consequently, the auxiliary Hilbert space decomposes into symmetric and antisymmetric subspaces under the action of \(R\), and the auxiliary Hamiltonian is unitarily equivalent to a direct sum of an isolated system Hamiltonian and the original interacting system, with the proper rescaling of the coupling operator.  

For simplicity, we retain the auxiliary basis, at the cost of working with replicated blocks rather than the corresponding direct-sums.

\subsection{Stationary-State Equivalence\label{Sec:EquEqv}}

 The boundary operator of Eqs.~\eqref{Eq:nresonantselectionBC} and~\eqref{Eq:VCPdefME1}  induces the respective decompositions
\begin{align}
N_{\rm rg}^{\rm aux}
&=
\partial N_{\rm rg}^{\rm aux}
+
(1-\partial)N_{\rm rg}^{\rm aux}\\
V^{\rm aux}&=
\partial V^{\rm aux}
+
(1-\partial)V^{\rm aux}.
\end{align}
 The auxiliary latent-resonance generator remains defined by selecting strictly vanishing Bohr frequencies.

We first examine the block-structure of the latent-resonance and boundary nonresonant superoperators,
\[
\tilde L_{\rm rg}^{\rm aux}
=
\begin{pmatrix}
\tilde  L_{\rm rg}^{\rm a} & \tilde  L_{\rm rg}^{\rm a}\\
\tilde  L_{\rm rg}^{\rm a} & \tilde  L_{\rm rg}^{\rm a}
\end{pmatrix},
\qquad
\partial  \tilde N_{\rm rg}^{\rm aux}
=
\begin{pmatrix}
\partial \tilde N_{\rm rg}^{\rm a} &
\partial \tilde N_{\rm rg}^{\rm a}\\
\partial \tilde N_{\rm rg}^{\rm a} &
\partial \tilde  N_{\rm rg}^{\rm a}
\end{pmatrix}.
\]
The tilde indicates that the diagonal matrix elements
\((i,i)\), \((i+N,i)\), and \((i,i+N)\) are omitted.
The resulting \(N\times N\) generators
\(L_{\rm rg}^{\rm a}\) and \(\partial N_{\rm rg}^{\rm a}\)
are completed by restoring the diagonal entries so that
each column sums to zero.

We compute \(L_{\rm rg}^{\rm a}\) and
\(\partial N_{\rm rg}^{\rm a}\) analytically.
For the latter, an arbitrarily small \(\epsilon\) is introduced in
Eq.~\eqref{Eq:nresonantselectionBC}, and the limit
\(\epsilon\to0\) is taken only after the relevant terms have
been selected. The derivation closely parallels that of
\(L_{\rm rg}\), with additional details given in
Appendix~\ref{Sec:LRGB}.

The final result is the stationary-state-preserving identity, which identifies the two latent-resonance generators with the auxiliary-boundary contribution through the same stationary-state-preserving transformer 
$\mathcal{T}^{\rm a}$,
\begin{equation}
L_{rg}\xleftrightarrow{\;\mathcal T^{\rm a}\;}L_{\rm rg}^{\rm a}\xleftrightarrow{\;-2\mathcal T^{\rm a}\;} \partial N_{\rm rg}^{\rm a}.
\label{Eq:thrippleEqEq}
\end{equation}
This identity unravels the latent resonances of the physical system to actual degeneracies in the auxiliary system.

Interestingly, the transformer is composed exclusively
by the principal-value integrals,
\begin{align}
\mathcal T^{\rm a}_{nn,ii}
=
2A_{nn}^{\rm a}A_{ii}^{\rm a}|A_{ni}^{\rm a}|^2
\frac{\mathcal P}{\pi}
\int_{-\infty}^{\infty}d\omega\,
J_{\omega}
\frac{J_{\omega_{in}}-J_{\omega_{in}-\omega}}
{\omega^2}.
\end{align}
As shown in Appendix~\ref{Sec:LRGB}, \(\mathcal T^{\rm a}\) annihilates the zeroth-order stationary state and thus defines a stationary-state-preserving transformation.
Stationary state equivalence also holds between the full auxiliary generators  $L_{\rm rg}^{\rm aux}$ and  $\partial N_{\rm rg}^{\rm aux}$. 

\subsection{Triangle Identity and the Stationary State\label{Sec:triangle}}

The triangle identity reintroduces the virtual coherence
pathway as the final stationary-state-preserving
transformation and the endpoint of the reduction
hierarchy.

Applying the boundary TCL4 selection rule
Eq.~\eqref{Eq:LocalBoundaryLaw} to the auxiliary system
yields
\[
\partial N_{\rm rg}^{\rm aux} p^{(0,{\rm aux})}
=
\partial V^{\rm aux} p^{(0,{\rm aux})}.
\]
Combined with the stationary-state equivalence established in
Eq.~\eqref{Eq:thrippleEqEq},
this yields the triangle identity,
\begin{equation}
L_{\rm rg}^{\rm aux} p^{(0,{\rm aux})}
=
\partial N_{\rm rg}^{\rm aux} p^{(0,{\rm aux})}
=
\partial V^{\rm aux} p^{(0,{\rm aux})},
\end{equation}
which expresses the stationary-state
equivalence of the latent-resonance generator and the
boundary projections of the nonresonant generator and
the virtual coherence pathway.

Applying $\partial V^{\rm a}$ to
the population in Eq.~\eqref{Eq:VCPdef2} yields
\begin{equation}
\partial V^{\rm aux} p^{(0,{\rm aux})}
=
-L^{(2,{\rm aux})}c^{(2,{\rm aux})}_{\epsilon},
\qquad \epsilon\neq0,
\end{equation}
where \(c^{(2,{\rm aux})}_{\epsilon}\) denotes the vector of
nearly resonant stationary coherences
\(c^{(2,{\rm aux})}_{n,n \pm N}\), selected by the projector
\(\partial\) through the definition of \(\partial V\),
Eq.~\eqref{Eq:VCPdefME1}. 
Because these coherences admit analytic
continuation to zero detuning~\cite{Thingna}, the
contraction \(\partial V^{\rm aux} p^{(0,{\rm aux})}\)
possesses a well-defined limit as \(\epsilon\to0\).

The triangle identity transforms the auxiliary stationary equation,
\[
L^{(2,{\rm aux})} p^{(2,{\rm aux})}
=
- L_{\rm rg}^{\rm aux} p^{(0,{\rm aux})},
\]
 to
\begin{equation}
L^{(2,{\rm aux})} p^{(2,{\rm aux})}
=
-\partial V^{\rm aux} p^{(0,{\rm aux})}
=
L^{(2,{\rm aux})}c^{(2,{\rm aux})}_\epsilon
\label{Eq:R2Eaux}
\end{equation}

Taking the limit \(\epsilon\to0\) therefore yields
\begin{equation}
p^{(2,{\rm aux})}
=
\lim_{\epsilon\to0}c^{(2,{\rm aux})}_\epsilon.
\label{Eq:analcontdd}
\end{equation}
Thus, the analytic continuation of the nearly resonant stationary
coherences yields the corresponding second-order
stationary population corrections in the auxiliary system. This demonstrates that the stationary-state corrections are already encoded in the organization of the second-order coherence dynamics.

The R2E$_2$ condition \(q^{(2, \rm{aux})}=p^{(2,{\rm aux})}\) follows
immediately, since the stationary and equilibrium
coherences coincide to second order~\cite{Fleming}. Under the convention of Eq.~\eqref{Eq:Aaux}, the pullback from the auxiliary to the physical system gives \begin{equation}
p^{(2)}=\sqrt{2}p^{(2,aux)}
\end{equation}

A natural question is how the virtual coherence pathway
should be interpreted. The intermediate coherences are never
occupied. However, in the limit of vanishing detuning
between the auxiliary levels, the corresponding
coherence acquires an effective population-like
character, since the timescale associated with its phase
oscillation becomes longer than all other relevant
dynamical timescales. The virtual process may
therefore be viewed as an effective transfer through a
quasi-static intermediate state: one population is
transferred to another while virtually occupying the
nearly stationary coherence.

In summary, stationary-state-preserving transformations progressively strip away the asymptotically redundant TCL4 structure until the correction equation is expressed entirely in terms of the virtual coherence pathway contraction.  Put simply, the virtual-coherence pathway emerges as the final stationary-state-preserving realization of the TCL4 population generator.


\section{Canonical Closures of the Redfield Equation}

We now derive canonically consistent closures of the
Redfield equation in the sense of Becker et al.~\cite{becker2022canonically}, which
were intended to steer the dynamics toward the
mean-force Gibbs state. In the present framework,
however, the closures emerge from successive
stationary-state-preserving simplifications of the TCL4
generator. The resulting
master equations retain the practicality of the Redfield
equation and the accuracy of its coherence dynamics,
while extending the same order of accuracy to the
population dynamics.

As noted in Ref.~\cite{becker2022canonically}, canonically consistent closures are not unique. In the present work, this nonuniqueness arises differently. The reduction of the TCL4 stationary-state equation generates a family of stationary-state equivalent representations connected by stationary-state-preserving transformations.

The closure problem can be stated as follows. We seek a
generator
\begin{equation}
L=L^{(0)}+\lambda^2L^{(2)}+\lambda^4\bar L^{(4)},
\end{equation}
where
\begin{equation}
\bar L^{(4)}
\xleftrightarrow{\;\mathcal T\;}
L^{(4)} .
\end{equation}
The stationary-state population correction then satisfies
\begin{equation}
L^{(2)}p^{(2)}
=
(-\bar L^{(4)}+V)p^{(0)},
\label{Eq:STbarV}
\end{equation}
without requiring the explicit TCL4 generator.
Since the correction acts entirely within the population
subspace, the lift of \(\bar L^{(4)}\) to the full Liouville
space is straightforward.

The stationary-state-preserving transformations
constructed above naturally lead to two closure paths.
The first retains an explicit generator representation,
\begin{equation}
\bar L^{(4)}
=
L_{\rm rg}+V.
\label{Eq:LrgClosure}
\end{equation}
Substituting Eq.~\eqref{Eq:LrgClosure} into
Eq.~\eqref{Eq:STbarV}, the virtual-coherence pathway
cancels identically, leaving the latent-resonance
generator alone. The stationary-state correction is then
recovered through Eq.~\eqref{Eq:ReducedAST}.

The second path proceeds only at the level of the
stationary-state contraction,
\begin{equation}
\bar L^{(4)}p^{(0)}
=
\partial V\,p^{(0)}
+
Vp^{(0)}.
\label{Eq:Vcont}
\end{equation}
Substituting into
Eq.~\eqref{Eq:STbarV} yields the same stationary-state
population correction as the faithful TCL4 generator by
virtue of the stationary-state equivalences established
above. Equation~\eqref{Eq:Vcont} is defined only after
contraction with \(p^{(0)}\), since \(\partial V\) is
not a finite generator prior to contraction.

The virtual-coherence generator \(V\) can have a large
norm. Once the contraction \(Vp^{(0)}\) has been
isolated, however, the generator \(V\) becomes
stationary-state redundant. We therefore replace the
\(V\) contribution in both closure paths by the
inhomogeneous term \(Vp^{(0)}\), preserving the
stationary-state correction while minimizing the
additional perturbation of the Redfield dynamics.

Thus, the first closure uses the latent-resonance generator,
\begin{equation}
L_{\rm cl}
=
L^{(0)}
+
\lambda^2L^{(2)}
+
\lambda^4
L^{\rm std},
\end{equation}
where
\[
(L^{\rm std})_{nn,ii}
=
L^{\rm std}_{nn,ii},
\qquad
(L^{\rm std})_{nm,ij}=0
\qquad\text{otherwise}.
\]
The corresponding master equation is
\begin{equation}
\frac{d\rho}{dt}
=
L_{\rm cl}\rho
+
\lambda^4\,\mathrm{diag}(V p^{(0)}).
\label{Eq:Lstdclosure}
\end{equation}

The second closure is the most compressed
representative of the second-order stationary-state-preserving class. It consists of the Redfield
generator $L_{\rm Re}=L^{(0)}+\lambda^2 L^{(2)}$, supplemented by the diagonal inhomogeneity
\begin{equation}
\label{Eq:diagIN}
\frac{d\rho}{dt}
=
L_{\rm Re}\rho
+
\lambda^4\,\mathrm{diag}(\bar V p^{(0)}),
\end{equation}
which is the main result, Eq.~\eqref{Eq:MainResult}.
Here $\bar V$ is the full virtual coherence pathway, with the resonant contributions to the contraction incorporated through the
analytic continuation of the  stationary coherences.


Unexpectedly, the endpoint of the reduction hierarchy is
a Redfield equation supplemented only by a single
inhomogeneous correction term constructed entirely from
the Redfield generator.

\begin{figure}[t]
\centering
\includegraphics[width=\columnwidth]{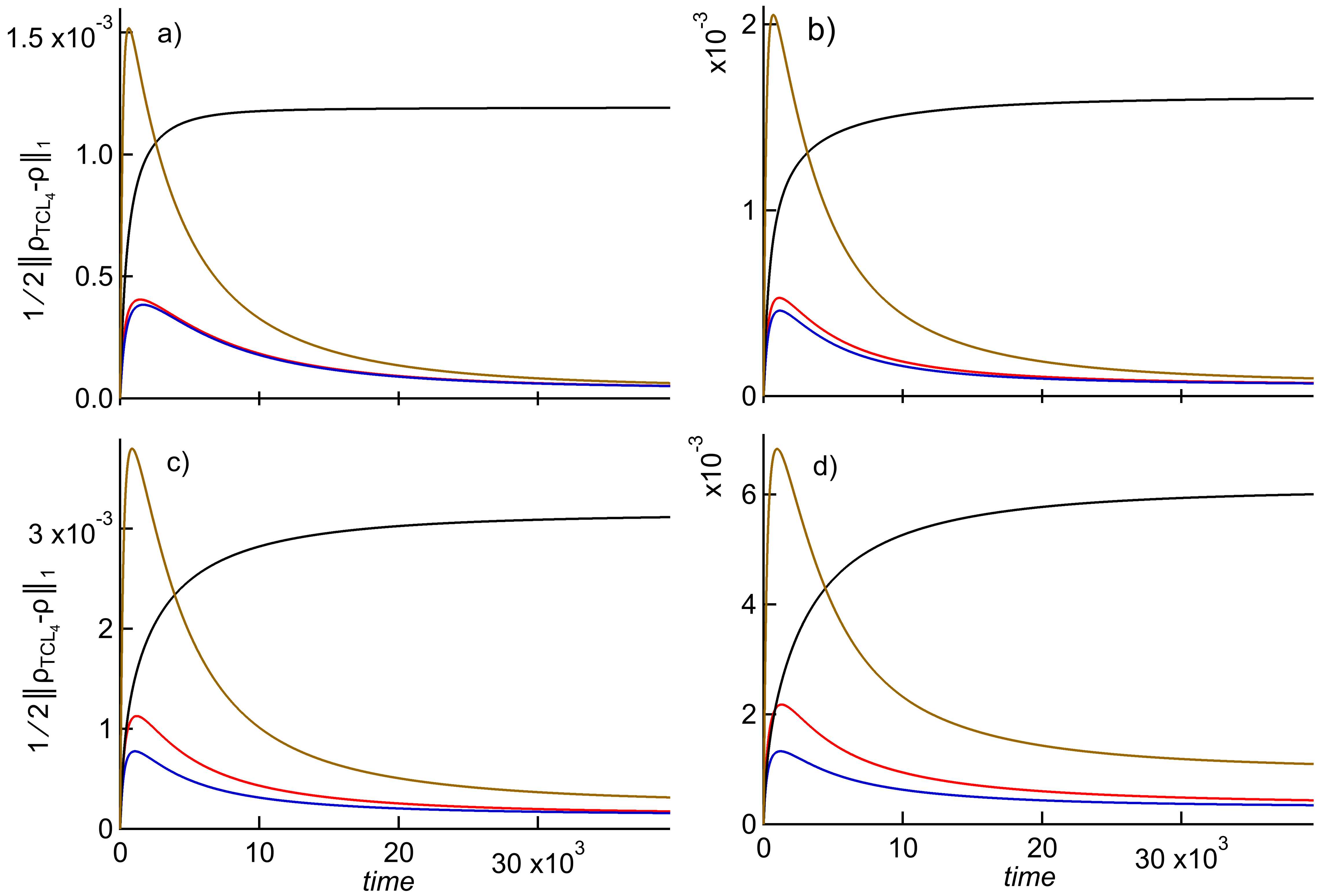}
\caption{
\textbf{Three canonically consistent closures of the Redfield equation.}
Time evolution of the trace distance between the TCL4
solution and the solutions of four master equations
for a four-level system. The black, red, blue, and brown 
curves correspond to the Redfield equation, the
latent-resonance closure, the 
virtual-coherence closure, and the expanded virtual-coherence closure, respectively. The dephasing parameter \(\xi=0,1,2,3\) in
panels (a)--(d), respectively.
\(\lambda^2=1.25\times10^{-3}\),
\(k_B T=0.26\).
The remaining parameters are given in
Appendix~\ref{App:SelectionRule}, Sample 1.}
\label{Fig:ClosureTraceDistance}
\end{figure}

We benchmark the closures using the trace distance
\[
D(\rho(t),\rho_{\rm TCL4}(t))
=
\frac12
\|\rho(t)-\rho_{\rm TCL4}(t)\|_1,
\]
where $\rho(t)$ is the solution of the Redfield equation with and without the closure.

We consider a four-level system whose energy spectrum is drawn from the center of the Gaussian unitary ensemble (GUE), scaled to have mean level spacing 1/3. Sampling from the center of the spectrum reproduces the universal Wigner--Dyson level-spacing statistics characteristic of the GUE. The system coupling operator is independently sampled from the ensemble of 4×4 GUE matrices and normalized to Frobenius norm 0.985. The diagonal matrix elements are multiplied by a parameter $\xi$, which controls the relative strength of diagonal coupling terms and the associated pure dephasing. We generate  $10^4$
 independent realizations by varying both the energy spectrum and the coupling operator. For each realization, the corresponding generators are constructed and the master equations are solved starting from the equal superposition state
 \[(\vert 1\rangle+\vert 2\rangle+\vert 3\rangle+\vert 4\rangle)/2.\]. 

To assess how the norm of the correction affects
accuracy, we introduce a third Redfield closure in which
the diagonal inhomogeneity in Eq.~\eqref{Eq:diagIN} is partially replaced by a
generator. The resulting master equation has the form of
Eq.~\eqref{Eq:Lstdclosure}, but with \(L_{\rm cl}\)
replaced by
\[
L_{\rm cl}'
=
L^{(2)}(\mathcal L+\mathcal L').
\]
Here, \(\mathcal L\) is a Pauli generator with
off-diagonal matrix elements
\[
\mathcal L_{nn,ii}
=
|A_{ni}|^2 S'_{in},
\]
while \(\mathcal L'\) is a diagonal matrix with elements
\begin{equation}
\mathcal L'_{nn,ii}
=
\beta \delta_{ni}
\sum_{k=1}^N
|A_{nk}|^2S_{nk}.
\end{equation}
This closure preserves the stationary state and should be viewed as intermediate between
the most compressed virtual-coherence closure of Eq.~\eqref{Eq:diagIN} and the
canonically consistent master equation of
Ref.~\cite{becker2022canonically}. 

At $\xi=0$, the average correction norms are \(0.89\), \(0.50\),
\(0.20\), and \(2.90\), respectively, for the full
\(L^{(4)}\) generator including coherence blocks, the
latent-resonance closure, the virtual-coherence closure,
and the expanded virtual-coherence closure.
Figure~\ref{Fig:ClosureTraceDistance} compares the average trace distance for the three closures. While individual realizations fluctuate, the ensemble averages reveal a clear trend: the virtual-coherence closure of Eq.~\eqref{Eq:diagIN}, which has the smallest correction norm, also exhibits the smallest transient error. The latent-resonance closure, despite retaining additional TCL4 information, performs less accurately on average. The expanded virtual-coherence closure further increases the correction norm and correspondingly degrades the transient accuracy. In all cases, however, the long-time behavior is substantially improved relative to the Redfield equation.

These results indicate that the norm of the correction is an important predictor of transient accuracy. For the benchmark considered here, increased compression is associated with both a smaller correction norm and improved dynamical fidelity.

\section{Discussion and Conclusions}

An important extension of the present framework is to
exact and near degeneracies. In closed quantum systems,
degenerate perturbation theory is essential and often
leads to qualitatively new physics. An analogous theory
for open-system equilibration remains largely
unexplored. Extending the present technique to
near-degenerate open quantum systems is therefore a
natural direction for future work.

The second direction concerns nonequilibrium boundary-driven quantum systems~\cite{RevModPhys.94.045006}. The present work used return to equilibrium as a benchmark for unraveling asymptotic simplifications.  A
similar chain of simplifications may exist for
nonequilibrium stationary states where such a benchmark does not exist. The Crowder's form of
the TCL4 generator is already applicable in such
settings~\cite{Crowder}, suggesting that
stationary-state-preserving transformations may possess
nonequilibrium analogues that preserve the stationary
state while reducing complexity. If so, the framework
developed here may provide a practical route toward
nonequilibrium stationary-state corrections without
resorting to complete TCL4 or numerical treatments.

In conclusion, we removed a large asymptotic redundancy
from the TCL4 generator, revealing an unexpected economy
hidden within its stationary limit. The resulting
reduction hierarchy isolates the stationary-state-carrying
content of TCL4 and compresses it into progressively
simpler stationary-state-preserving representations,
culminating in the virtual coherence pathway.

The hierarchy resolves the longstanding
stationary-state problem of the Redfield master equation,
compressing the required TCL4 correction into a simple
inhomogeneous diagonal term while restoring second-order perturbative accuracy. Remarkably, in the examples
considered here, the most compressed closure also provides the highest accuracy. Thus, increased
compression does not degrade fidelity; if anything, the
opposite trend is observed.

More broadly, the results suggest that the apparent
complexity of higher-order open-system dynamics may
reflect substantial asymptotic redundancy, with much
simpler structures emerging once attention is restricted
to stationary-state properties.

\section*{Acknowledgments}
The author thanks Prem Kumar and Juzar Thingna for fruitful and stimulating discussions. This work was supported by the David and Lucile Packard Foundation and by the School of Physics at the Georgia Institute of Technology through a seed grant.

\appendix
\section{Verification of the Selection Rule\label{App:SelectionRule}}

We begin by proving the selection rule Eq.~\eqref{Eq:SelectionRule} explicitly for the qubit
Hamiltonian
\[
H_S=\frac{\Delta}{2}\sigma_z,
\qquad
A=\frac12(\sigma_x+\xi\sigma_z).
\]

\subsection{Virtual coherence pathway}
The second order coherence correction~\eqref{Eq:c2} is
\begin{align}
\rho_{12}^{(2)}
&=
-\frac{i}{\Delta}
\left[
L^{(2)}_{12,11}p_1^{(0)}
+
L^{(2)}_{12,22}p_2^{(0)}
\right]
\nonumber\\
&=
-\frac{i}{\Delta}
\Big[
2A_{12}A_{11}
(\Gamma_\Delta^\ast+iS_0)p_1^{(0)}
\nonumber\\
&\qquad\qquad
+
2A_{12}A_{11}
(-\Gamma_{-\Delta}+iS_0)p_2^{(0)}
\Big]
\nonumber\\
&=
\frac{2A_{12}A_{11}}{\Delta}
\left(
S_0
-
p_1^{(0)}S_\Delta
-
p_2^{(0)}S_{-\Delta}
\right),
\end{align}
where the dissipative contributions cancel by detailed
balance. These are precisely the perturbative coherence
corrections appearing in the mean-force Gibbs state,
whose resonant continuation generates the equilibrium
population correction~\cite{Thingna,cresser2021weak}.

The corresponding virtual coherence pathway~\eqref{Eq:VCPdef} contraction is
\begin{equation}
{\sum_{lm}}'
L^{(2)}_{11,lm}\rho^{(2)}_{lm}
=
8|A_{12}|^2A_{11}^2J_0
\frac{
S_0
-
p_1^{(0)}S_\Delta
-
p_2^{(0)}S_{-\Delta}
}{
\Delta
}.\label{Eq:VCP}
\end{equation}
By trace preservation, the contribution to
the \(22\) population equation is equal and opposite, so
it suffices to consider the \(11\) component.

\subsection{Useful Identities}

The evaluation of Eq.~\eqref{Eq:RtildeFreq}
requires the Plemelj--Sokhotski identity
\begin{equation}
\frac{1}{x-i0^+}
=
i\pi\delta(x)
+
\mathcal P\frac{1}{x},
\end{equation}
where $\mathcal P$ denotes the principal value.

This converts the real part of Eq.~\eqref{Eq:RtildeFreq} to
\begin{align}
{\rm Re}\,\tilde{R}
&=-J_{\omega_2}\frac{S_{\omega_1}-S_{-\omega_2-\omega_3}}{\Omega}\nonumber\\
&+\frac{\mathcal P}{\pi}
\int_{-\infty}^{\infty}
d\omega\,
J_\omega
\frac{
J_{\omega_1}
-
J_{-\omega_3-\omega}
}{
(\omega-\omega_2)
(\omega_1+\omega_3+\omega)
}.
\label{Eq:ReR}
\end{align}
Eq.~\eqref{Eq:RtildeTime} shows the transposition symmetry
\begin{equation}
\tilde R_{\omega_1,\omega_2,\omega_3}
=
\tilde R_{\omega_2,\omega_1,\omega_3},
\label{Eq:Rtildesymmetry}
\end{equation}
The Kramers-Kronig relation connects the real and imaginary parts of $\Gamma_\omega$,
\begin{equation}
S_\omega=\frac{\mathcal P}{\pi}\int d\omega'\,\frac{J_{\omega'}}{\omega-\omega'}.
\end{equation} 
Inserting into Eq.~\ref{Eq:ReR} yields
\begin{align}
\operatorname{Re}\tilde{R}
&=
-\frac{
J_{\omega_2}
\left(
S_{\omega_1}
-
S_{-\omega_2-\omega_3}
\right)
+
J_{\omega_1}
\left(
S_{\omega_2}
-
S_{-\omega_1-\omega_3}
\right)
}{
\omega_1+\omega_2+\omega_3
}
\nonumber\\
&\quad
-
\frac{\mathcal P}{\pi}
\int d\omega\,
\frac{
J_\omega J_{-\omega_3-\omega}
}{
(\omega-\omega_2)
(\omega_1+\omega_3+\omega)
}
\label{Eq:ReR1}.
\end{align}

Another KMS identity is
\begin{equation}
\frac{\mathcal P}{\pi}
\int d\omega'\,
J_{\omega'}
\frac{
e^{-\beta(\omega'-\omega)}
-
1
}{
(\omega-\omega')^2
}
=
S'(\omega)
-
e^{\beta\omega}
S'(-\omega).
\label{Eq:KMSS}
\end{equation}

\subsection{Nonresonant population generator $N_{rg}$}

We evaluate $N_{rg}$ directly from Eq.~\eqref{Eq:nresonantselection}. We suppress the subscript $rg$ and the superscipt $(0)$ in Gibbs populations. Counting all the terms where $\Omega=0$, the \(11\) component of the contraction
takes the form
\begin{align}
&\sum N_{11,kk}p_k=p_1\Big[2A_{12}A_{21}A_{11}^2\,F_{11,11,21}
+\vert A_{12}\vert^4\,F_{21,21,22} \nonumber\\
&-2A_{12}A_{21}A_{11}^2C_{11,11,21}
+ (A_{12}A_{21})^2\,C_{12,12,11}\nonumber\\
&+\vert A_{12}\vert^4[R_{12,12,11}-R_{21,21,11}]-4\vert A_{12}\vert^2 A_{11}^2 R_{11,11,21}\Big]\nonumber\\
&+\Big[- \vert A_{12}\vert^4\,F_{12,12,11}
- 2A_{12}A_{21}A_{11}^2\,F_{22,22,12}\nonumber\\
&-(A_{12}A_{21})^2\,C_{21,21,22}
+2A_{11}^2A_{12}A_{21}\,C_{11,22,12} \nonumber\\
&+4A_{11}^2\vert A_{12}\vert^2\,R_{11,11,12}+
\vert A_{12}\vert^4
[
R_{12,12,22}-R_{21,21,22}
\Big]p_2\nonumber\\
&+\text{c.c.}
\label{Eq:nrfull}
\end{align}
We first collect the the nonresonant frequency-denominator
terms in \(F\), \(C\), and \(R\) in Eqs.~\eqref{Eq:TCLF}-\eqref{Eq:TCLC}, leaving the
\(\tilde R\) contribution aside. This gives after cancellations
\begin{align}
&\sum N_{11,kk} p_k=p_1\Big[
-i\vert A_{12}\vert^4\frac{(\Gamma_\Delta-\Gamma_{-\Delta})^2}{2\Delta} \nonumber\\
&+4i\vert A_{12}\vert^2A_{11}^2J_0\frac{\Gamma_\Delta-\Gamma_0}{\Delta} \Big]+p_2\Big[
-i\vert A_{12}\vert^4\frac{(\Gamma_\Delta-\Gamma_{-\Delta})^2}{2\Delta}\nonumber\\
&
-4i\vert A_{12}\vert^2 A_{11}^2 J_0 \frac{\Gamma_0-\Gamma_{-\Delta}}{\Delta}\Big]+\text{c.c.}.
\end{align}

Adding the complex conjugate reduces all algebraic terms to
\begin{align}
\sum_k N_{11,kk} p_k&=
\vert A_{12}\vert^4\text{Im}\frac{(\Gamma_\Delta-\Gamma_{-\Delta})^2}{\Delta}\nonumber\\
&+8\vert A_{12}\vert^2A_{11}^2J_0\frac{S_0-p_1 S_\Delta-p_2 S_{-\Delta}}{\Delta}
\label{Eq:nralgebraic}
\end{align}

It remains to add the integral terms that depend on \(\tilde R\).
The remaining identities to prove are
\begin{align}
2\operatorname{Re}
\left(
\tilde R_{21,21,11}
-
\tilde R_{12,12,11}
\right)
&=
\operatorname{Im}
\frac{
(\Gamma_\Delta-\Gamma_{-\Delta})^2
}{
\Delta
},
\label{Eq:RIdentity1}
\end{align}
\begin{align}
\operatorname{Re}
\Big(
p_1\tilde R_{11,11,21}-&
p_2\tilde R_{11,11,12}\Big)\nonumber\\
&=2J_0
\frac{
S_0-p_1S_\Delta-p_2S_{-\Delta}
}{
\Delta
}.
\label{Eq:RIdentity2}
\end{align}

We  first prove Eq.~\eqref{Eq:RIdentity1}. For the two terms entering
\(\tilde R_{21,21,11}-\tilde R_{12,12,11}\), the principal-value
integrals from Eq.~\eqref{Eq:ReR} are identical immediately and cancel. The remaining
nonresonant algebraic terms give
\begin{align}
&2\operatorname{Re}
\left(
\tilde R_{21,21,11}
-
\tilde R_{12,12,11}
\right)\nonumber\\
&=
2\left[
-\frac{
2J_{-\Delta}(S_{-\Delta}-S_\Delta)
}{
-2\Delta
}+
\frac{
2J_{\Delta}(S_{\Delta}-S_{-\Delta})
}{
2\Delta
}
\right]
\nonumber\\
&=
2\frac{
(J_\Delta-J_{-\Delta})(S_\Delta-S_{-\Delta})
}{
\Delta
}.
\label{Eq:RIdentity1Proof}
\end{align}
Using
\[
\Gamma_\Delta-\Gamma_{-\Delta}
=
(J_\Delta-J_{-\Delta})
+
i(S_\Delta-S_{-\Delta}),
\]
one has
\[
\operatorname{Im}
\frac{
(\Gamma_\Delta-\Gamma_{-\Delta})^2
}{
\Delta
}
=
2\frac{
(J_\Delta-J_{-\Delta})(S_\Delta-S_{-\Delta})
}{
\Delta
},
\]
which proves Eq.~\eqref{Eq:RIdentity1}.

We next prove Eq.~\eqref{Eq:RIdentity2}. The algebraic part of Eq.~\eqref{Eq:ReR} yields
\begin{align}
&\operatorname{Re}
\left(
p_1\tilde R_{11,11,21}
-
p_2\tilde R_{11,11,12}
\right)_{\rm alg}
\nonumber\\
&\qquad
=
-p_1
\frac{
2J_0(S_0-S_\Delta)
}{
-\Delta
}
+
p_2
\frac{
2J_0(S_0-S_{-\Delta})
}{
\Delta
}
\nonumber\\
&\qquad
=
2J_0
\frac{
S_0-p_1S_\Delta-p_2S_{-\Delta}
}{
\Delta
}.
\label{Eq:RIdentity2Alg}
\end{align}
It remains to show that the principal-value integrals from Eq.~\eqref{Eq:ReR} cancel. The two
principal-value contributions are
\begin{align}
p_1
\mathcal P\int d\omega\,
\frac{
J_\omega J_{\Delta-\omega}
}{
\omega(\omega-\Delta)
}
\end{align}
and
\begin{align}
p_2
\mathcal P\int d\omega\,
\frac{
J_\omega J_{-\Delta-\omega}
}{
\omega(\omega+\Delta)
}.
\end{align}
Using the KMS relation \(J_{-\omega}=e^{-\beta\omega}J_\omega\) and
\(p_2=e^{-\beta\Delta}p_1\), one finds
\begin{align}
p_1
\mathcal P\int d\omega\,
\frac{
J_\omega J_{\Delta-\omega}
}{
\omega(\omega-\Delta)
}
&=
p_1
\mathcal P\int d\omega\,
\frac{
e^{\beta\omega}J_{-\omega}\,
e^{\beta(\Delta-\omega)}J_{\omega-\Delta}
}{
\omega(\omega-\Delta)
}
\nonumber\\
&=
p_2
\mathcal P\int d\omega\,
\frac{
J_{-\omega}J_{\omega-\Delta}
}{
\omega(\omega-\Delta)
}.
\end{align}
Changing variables \(\omega\mapsto-\omega\) gives
\begin{align}
p_1
\mathcal P\int d\omega\,
\frac{
J_\omega J_{\Delta-\omega}
}{
\omega(\omega-\Delta)
}
&=
p_2
\mathcal P\int d\omega\,
\frac{
J_{\omega}J_{-\Delta-\omega}
}{
\omega(\omega+\Delta)
}.
\end{align}
Therefore the principal-values of $\tilde R$ in
the left-hand side of Eq.~\eqref{Eq:RIdentity2} cancel, leaving only the algebraic term in Eq.~\eqref{Eq:RIdentity2Alg} and Eq.~\eqref{Eq:RIdentity2}
follows.

Combining Eqs.~\eqref{Eq:RIdentity1} and \eqref{Eq:RIdentity2}
with the \(\tilde R\) terms in Eq.~\eqref{Eq:nrfull}, and adding
the contraction given by Eq.~\eqref{Eq:VCP} yields the cancellation
\begin{equation}
\sum_k N_{11,kk}p_k+
{\sum_{lm}}'
L_{11,lm}^{(2)}\rho_{lm}^{(2)}=0.
\label{Eq:QubitSelection11}
\end{equation}

By trace preservation, the \(22\) component is equal and
opposite. The cohernce term is related to VCP via Eqs.~\eqref{Eq:VCPdef} and~\ref{Eq:c2} in main text and the selection rule Eq.~\eqref{Eq:SelectionRule} is verified,
\begin{equation}
\sum_k N_{11,kk}p_k=\sum_k V_{11,kk}p_k.
\end{equation}

\subsection{Numerical demonstrations beyond a qubit\label{app:numerics}}
Generalizing the qubit proof to arbitrary
finite-dimensional systems is challenging because the
selection rule originates from the three-frequency
combinatorics of the TCL4 generator. A general proof of the
selection rule for arbitrary \(N\) remains unavailable.

As a non-qubit check, we sampled an \(N=4\)
Hamiltonian \(H_S\) and coupling operator \(A\) from the
Gaussian unitary ensemble and evaluated the
selection-rule residual
\[
\epsilon
=
\frac{
\left\|
(N_{\rm rg}-V)p^{(0)}
\right\|
}{
\left\|
N_{\rm rg}p^{(0)}
\right\|
}
\]
for a Drude--Lorentz spectral density,
\begin{equation}
    J_\omega=2\pi\frac{\omega}{1+(\frac{\omega}{\omega_c})^2}.
\end{equation} and over several
temperatures. 

For the representative benchmark Example~1,
\begin{widetext}
\begin{align}
A =
\begin{pmatrix}
0.560 & 0.196 + 0.326 i & -0.116 + 0.054 i & 0.092 + 0.238 i \\
0.196 - 0.326 i & -0.625 & 0.083 + 0.327 i & 0.013 - 0.128 i \\
-0.116 - 0.054 i & 0.083 - 0.327 i & 0.617 & 0.064 - 0.212 i \\
0.092 - 0.238 i & 0.013 + 0.128 i & 0.064 + 0.212 i & 0.094
\end{pmatrix},
\end{align}
\end{widetext}
with
\begin{align}
E =
(
0.940,
0.779,
-0.434,
-1.041)^T
\end{align}
and parameters
$\omega_c = 4,
kT = 0.26,
dt = 10^{-6},
t = 8.39,$
where \(dt\) and \(t\) denote the time step and integration
time used for evaluating \(\tilde R\). The dominant numerical
error originates from the finite time step, quantified by
\[
\frac{\|\tilde R_{t/2,dt}-\tilde R_{t/2,dt/2}\|}
{\|\tilde R_{t/2,dt/2}\|}
=
2.51\times10^{-12}.
\]
Here, \(\tilde R_{x,y}\) denotes the numerical result
obtained using integration time \(x\) and time step \(y\).
Other numerical errors are smaller by an order of
magnitude. Numerically we obtain
\[
\epsilon = 1.13\times10^{-12},
\]
which is consistent with the integration accuracy.

 Example~2 differs from
Example~1 only by moving the upper level to a near
degeneracy,
\[
E'_1 = E_2 + 10^{-7}.
\]
The norm of the virtual coherence pathway increases by
approximately \(2\times10^6\) due to the near-vanishing
energy denominator. Nevertheless, the selection-rule
residual remains small,
\[
\epsilon \approx 2\times10^{-9}.
\]
Thus, the contraction with the Gibbs state continues to
enforce the cancellation even when the degeneracy is closely approached.

\subsubsection{Decomposition~\label{Ap:stratificationexample}}

We next compare the contractions of the virtual
coherence pathway and \(N_{\rm rg}\) on the
near-resonant pair \((34)\) and in the nonresonant
interior, to show how the selection rule
stratifies into two independent identities Eqs.~\eqref{Eq:LocalBoundaryLaw} and~\eqref{Eq:LocalBoundaryComplement}. We set $E(3)=E(4)+1\times 10^{-6}$.

$\partial_{34}N_{\rm rg}$ is obtained by restricting the sum
in Eq.~\eqref{Eq:nresonantselection} to terms satisfying
$0<|\Omega|<10^{-5}$, thereby isolating the contribution
associated with the boundary $\omega_{34}\to0$, while
$(1-\partial_{34})N_{\rm rg}$ denotes the complementary
remainder. Similarly, $\partial_{34}V$ is obtained by
restricting the sum
\[
V_{ll,ii}
=
i{\sum_{ab}}'
L_{ll,ab}^{(2)}
\frac{1}{\omega_{ab}}
L_{ab,ii}^{(2)},
\]
to terms satisfying
$0<|\omega_{ab}|<10^{-5}$, while
$(1-\partial_{34})V$ denotes the complementary remainder.

On the boundary, we obtain
\begin{align}
\frac{
\left\|
\partial_{34}
\big(
N_{rg}-V
\big)p^{(0)}
\right\|
}{
\left\|
\partial_{34}N_{rg}p^{(0)}
\right\|
}
=
9.7\times10^{-8},
\end{align}
and in the interior,
\begin{align}
\frac{
\left\|
(1-\partial_{34})
\big(
N_{rg}-V
\big)p^{(0)}
\right\|
}{
\left\|
\partial_{34}N_{rg}p^{(0)}
\right\|
}
=
1.1\times10^{-12},
\end{align}
while the vectors $\partial_{34}
N_{rg}p^{0}$ and $(1-\partial_{34})
N_{rg}p^{0}$ remain linearly independent, forming an angle of \(1.92\) radians.

Thus, the selection rule decomposes into independent
cancellations on the near-resonant boundary and in the
nonresonant interior. The dominant numerical error remains
concentrated on the boundary, where the norm of the virtual
coherence pathway is strongly enhanced, while the interior
cancellation remains comparable to the near uniform-gap
case.

We examined several additional approaches to degeneracy,
including multiple near-degenerate pairs. In all cases the selection rule remained
accurate to numerical precision, and the boundary and
interior contributions satisfied the decomposed identities
with comparable accuracy.

These results support both the generality of the selection
rule and its decomposition into independent boundary and
interior constraints. 

\section{Latent-Resonance and Boundary Generators \label{Sec:LRGB}}

We suppress the subscript $rg$ in $L_{rg}$  until the end of the derivation. Collecting the terms $\Omega=0$ in Eq.~\ref{Eq:TCL4Population}, we find, 
\begin{align}
&L_{nn,ii}\{A\}
=
|A_{ni}|^4
\Big[
 C_{ni,in,ii}
+ F_{in,ni,nn}
\Big]
\nonumber\\
&\quad
-
|A_{ni}|^4
\Big[
 R_{in,ni,ii}
- R_{ni,in,ii}
\Big]
\nonumber\\
&\quad+
|A_{ni}|^2
A_{ii}(A_{nn}-A_{ii})
\Big[
 C_{in,ii,ni}
- F_{in,ii,ni}
\Big]
\nonumber\\
&\quad-
|A_{ni}|^2(A_{nn}-A_{ii})^2
R_{in,ii,ni}
\nonumber\\
&\quad+
\sum_{\substack{k\neq n\\k\neq i}}
\Bigg\{
|A_{ni}|^2|A_{nk}|^2
\Big[
 C_{nk,in,ki}
- R_{in,nk,ki}
+ R_{nk,in,ki}
\Big]
\nonumber\\
&\qquad\qquad
-
|A_{ik}|^2|A_{nk}|^2
\Big[
 C_{kn,ik,ni}
+ R_{kn,ik,ni}
\Big]
\nonumber\\
&\qquad\qquad
+
|A_{ni}|^2|A_{ik}|^2
F_{in,ki,nk}
\Bigg\}
+\text{c.c.}.\label{Eq:Lrg00}
\end{align}
The limit $\Omega\to 0$ in Eqs.~\eqref{Eq:TCLF}-\eqref{Eq:TCLC} yields latent-resonance derivatives
\begin{align}
F_{\omega_1\omega_2\omega_3}
&=
-i\,\Gamma^{T}_{\omega_2}
\Gamma'_{\omega_1},\label{Eq:TCLFd}
\\[8pt]
C_{\omega_1\omega_2\omega_3}
&=
-i\,\Gamma^{\star}_{\omega_2}
\Gamma'_{\omega_1},\label{Eq:TCLCd}
\\[8pt]
R_{\omega_1\omega_2\omega_3}
&=
\tilde{R}_{\omega_1\omega_2\omega_3}
-
i\,\Gamma_{\omega_2}
\Gamma'_{\omega_1},\label{Eq:TCLRd}
\end{align}
which are to be inserted in Eqs.~\eqref{Eq:Lrg00} and ~\eqref{Eq:Nrg00}.

The auxiliary boundary generator block \(\partial N_{\rm rg}^{\rm a}\) introduced in Sec.~\ref{Sec:EquEqv} is obtained from Eqs.~\eqref{Eq:resonantselection} and~\eqref{Eq:nresonantselectionBC}. We first assume arbitrarily small \(\epsilon\) to isolate the boundary-shell contributions and only subsequently take the limit \(\epsilon\to0\). The algebra closely parallels the derivation of \(L_{\rm rg}\), yielding
\begin{align}
(\partial N_{\rm rg}^{\rm a})_{nn,ii}
&=
L_{nn,ii}\{A^{\rm a}\}
+
\mathcal T_{nn,ii},
\end{align}
where
\begin{align}
\mathcal T_{nn,ii}
&=
A_{nn}^{\rm a}A_{ii}^{\rm a}|A_{ni}^{\rm a}|^2
\operatorname{Re}
\bigl(
C_{in,0,ni}
-
F_{in,0,ni}
+
2R_{in,0,ni}
\bigr)
\label{Eq:Nrg00}
\\
&=
2A_{nn}^{\rm a}A_{ii}^{\rm a}|A_{ni}^{\rm a}|^2
\,\mathcal U_{nn,ii},
\nonumber
\end{align}
with
\begin{equation}
\mathcal U_{nn,ii}
=
\frac{\mathcal P}{\pi}
\int_{-\infty}^{\infty}d\omega\,
J_\omega
\frac{J_{\omega_{in}}-J_{\omega_{in}-\omega}}
{\omega^2}.
\label{Eq:Ukernel}
\end{equation}
The principal-value \(\mathcal U\) is the remnant of Eq.~\eqref{Eq:ReR}, while all the derivative terms  cancel identically.

To establish that \(\mathcal T\) defines an stationary-state-preserving transformer, it suffices to show that \(\mathcal U\) annihilates the Gibbs pair currents,
\begin{equation}
\mathcal U_{nn,ii}p_i
-
\mathcal U_{ii,nn}p_n
=
0.
\label{Eq:Upair}
\end{equation}
Summation over \(i\), together with trace preservation, then recovers the stationary-state preserving condition, Eq.~\eqref{Eq:EqEquiv}.

Indeed,
\begin{align}
\mathcal U_{nn,ii}p_i
-&
\mathcal U_{ii,nn}p_n
=
\frac{\mathcal P}{\pi}
\int d\omega\,
\frac{J_\omega}{\omega^2}
\nonumber\\
&\times
\Big[
p_i\bigl(J_{in}-J_{\omega_{in}-\omega}\bigr)
-
p_n\bigl(J_{ni}-J_{\omega_{ni}-\omega}\bigr)
\Big]
\nonumber\\
&=
\frac{\mathcal P}{\pi}
\int d\omega\,
\frac{J_\omega}{\omega^2}
\,p_n
\Big[
-e^{-\beta\omega}
J_{\omega-\omega_{in}}
+
J_{\omega_{ni}-\omega}
\Big].\label{Eq:piarU}
\end{align}
In obtaining the second line we used the KMS identities
\begin{equation}
    \label{Eq:KMS00}
p_iJ_{in}=p_nJ_{ni},
\qquad
p_iJ_{\omega_{in}-\omega}
=
p_ne^{-\beta\omega}
J_{\omega-\omega_{in}}.
\end{equation}
Splitting the principal-value integral into two parts,
performing the change of variables \(\omega\to-\omega\)
in one contribution, and using the KMS relation once
more, the two terms cancel identically.

\subsection{Explicit Latent-Resonance Generator and its Standard Form}

Leaving $\tilde R$ aside, temporarily, 
the matrix element becomes
\begin{align}
&L_{nn,ii}=
|A_{ni}|^4
\Big[
-2iJ_{in}\Gamma'_{ni}
-i(\Gamma_{in}-\Gamma_{ni})\Gamma'_{in}
\Big]
\nonumber\\
&+
i|A_{ni}|^2
A_{ii}(A_{nn}-A_{ii})
2iS_0\Gamma'_{in}
\nonumber\\
&+
i|A_{ni}|^2
(A_{nn}-A_{ii})^2
\Gamma_0\Gamma'_{in}
\nonumber\\
&+
i\sum_{\substack{k\neq n\\k\neq i}}
\Bigg\{
|A_{ni}|^2|A_{nk}|^2
\Big[
-2J_{in}\Gamma'_{nk}
+
\Gamma_{nk}\Gamma'_{in}
\Big]
\nonumber\\
&\qquad
+
|A_{ik}|^2|A_{nk}|^2
2J_{ik}\Gamma'_{kn}
-
|A_{ni}|^2|A_{ik}|^2
\Gamma_{ik}\Gamma'_{in}
\Bigg\}
+\text{c.c.}.
\end{align}
Collecting terms further gives
\begin{align}
&L_{nn,ii}=
-i|A_{ni}|^2\Gamma'_{in}
\Big[
\sum_{k=1}^N
\Big(
|A_{ik}|^2\Gamma_{ik}
-
|A_{nk}|^2\Gamma_{nk}
\Big)
\nonumber\\
&\qquad
+
2J_0A_{ii}(A_{nn}-A_{ii})
\Big]
\label{Eq:FGRembed}\\
&-
2i|A_{ni}|^4
J_{in}\Gamma'_{ni}
\label{Eq:NONembed}\\
&+
2i
\sum_{\substack{k\neq n\\k\neq i}}
|A_{nk}|^2
\Big[
-|A_{ni}|^2
J_{in}\Gamma'_{nk}
+
|A_{ik}|^2
J_{ik}\Gamma'_{kn}
\Big]
+\text{c.c.}.
\label{Eq:StandardForm0}
\end{align}

The first term~\eqref{Eq:FGRembed} is a
renormalized Fermi-golden-rule (FGR) contribution together
with dephasing-induced spectral overlap broadening~\cite{Lampert2025}. The
remaining terms describe additional latent-resonances that cannot be embedded into a simple
FGR/spectral-overlap picture.
The doubly restricted sum ~\eqref{Eq:StandardForm0} vanishes identically in the
qubit case. 

We now return to the omitted resonant contribution
generated by \(\tilde R\),
\begin{align}
\tilde L_{nn,ii}
&=
-|A_{ni}|^2
(A_{nn}-A_{ii})^2
\tilde R_{in,ii,ni}
+\text{c.c.}
\label{Eq:LrgA1a}
\\
&\quad
-
\sum_{\substack{k\neq n\\k\neq i}}
|A_{ik}|^2|A_{nk}|^2
\tilde R_{kn,ik,ni}
+\text{c.c.}
\label{Eq:LrgA1b}
\end{align}

This contains principal-value integrals that are independent from the germs generating the mean-force Gibbs correction. We show that they can be eliminated through stationary-state-preserving transformations.

\subsection{Elimination of Principal Integrals\label{Ap:PIelimination}}

We now derive the standard form of the latent-resonance generator~\eqref{Eq:StandardForm} , which is devoid of principle integrals.

Due to the complex conjugates in the previous equations,
only the real part of \(\tilde R\) contributes. Substituting
the principal-value part of Eq.~\eqref{Eq:ReR} into
Eq.~\eqref{Eq:LrgA1a}, we find that the principal-value contribution reduces to the pair-current difference~\eqref{Eq:piarU}
\begin{equation}
-
|A_{ni}|^2
(A_{nn}-A_{ii})^2(
\mathcal U_{nn,ii}p_i-
\mathcal U_{ii,nn}p_n)=0.
\end{equation}
Thus Eq.~\eqref{Eq:LrgA1a} is stationary-state-preserving to 
\begin{align}
\tilde L_{nn,ii}
&=
|A_{ni}|^2
(A_{nn}-A_{ii})^2
J_0S'_{in}
+\text{c.c.}
\nonumber\\
&\quad
-
\sum_{\substack{k\neq n\\k\neq i}}
|A_{ik}|^2|A_{nk}|^2
\tilde R_{kn,ik,ni}
+\text{c.c.}.
\end{align}

The remaining principal-value contribution in the second line is eliminated next.
Define
\begin{equation}
C_{ni}
=
-
\sum_{\substack{k\neq n\\k\neq i}}
|A_{ik}|^2|A_{nk}|^2
\tilde R_{kn,ik,ni}
\end{equation}
and
\begin{equation}
\Psi_{ni}
=
C_{ni}p_i
-
C_{in}p_n.
\label{Eq:Psi_asym}
\end{equation}
Applying the transposition symmetry~\eqref{Eq:Rtildesymmetry},
we obtain two transposed representatives,
\begin{align}
\Psi_{ni}^1
&=
-
\sum_{\substack{k\neq n\\k\neq i}}
|A_{ik}|^2|A_{nk}|^2
\Big[
p_i\tilde R_{kn,ik,ni}
-
p_n\tilde R_{nk,ki,in}
\Big]
\label{Eq:Psi3}\\
\Psi_{ni}^2
&=
-
\sum_{\substack{k\neq n\\k\neq i}}
|A_{ik}|^2|A_{nk}|^2
\Big[
p_i\tilde R_{ik,kn,ni}
-
p_n\tilde R_{ki,nk,in}
\Big].
\end{align}

Evaluating the resonant part of the
Sokhotski--Plemelj decomposition~\eqref{Eq:ReR} of the two representatives
yields,
\begin{align}
\Psi_{ni}^{r1}
&=
-
\sum_{\substack{k\neq n\\k\neq i}}
|A_{ik}|^2|A_{nk}|^2
J_{kn}
\Big[
p_kS'(\omega_{ki})
-
p_iS'(\omega_{ik})
\Big],
\label{Eq:Pem1}
\\
\Psi_{ni}^{r2}
&=
-
\sum_{\substack{k\neq n\\k\neq i}}
|A_{ik}|^2|A_{nk}|^2
J_{ki}
\Big[
p_nS'(\omega_{nk})
-
p_kS'(\omega_{kn})
\Big].
\label{Eq:Pem2}
\end{align}

We next evaluate the principal-value contribution of the
first representative,
\begin{align}
\Psi_{ni}^{p1}
&=
-\frac{1}{\pi}
\sum_{\substack{k\neq n\\k\neq i}}
|A_{ik}|^2|A_{nk}|^2
\mathcal P
\int d\omega\,J_\omega
\nonumber\\
&\qquad\times
\Bigg[
p_i
\frac{
J_{\omega_{ik}}
-
J_{\omega_{in}-\omega}
}{
(\omega-\omega_{kn})^2
}
-
p_n
\frac{
J_{\omega_{ki}}
-
J_{\omega_{ni}-\omega}
}{
(\omega-\omega_{nk})^2
}
\Bigg].
\label{Eq:PRPR}
\end{align}

After the change of variables
\(
\omega\rightarrow-\omega
\)
and application of the KMS condition, all terms involving
the spectral-density continuation
\(J_{\omega_{ni}-\omega}\)
cancel identically. The remaining expression reduces to
\begin{align}
\Psi_{ni}^{p1}
&=
-\frac{p_n}{\pi}
\sum_{\substack{k\neq n\\k\neq i}}
|A_{ik}|^2|A_{nk}|^2
J_{ki}
\mathcal P
\int d\omega J_\omega
\frac{
e^{-\beta(\omega-\omega_{nk})}
-
1
}{
(\omega-\omega_{nk})^2
}.
\end{align}

Applying the KMS identity~\eqref{Eq:KMSS}, we obtain
\begin{align}
\Psi_{ni}^{p1}
&=
-\sum_{\substack{k\neq n\\k\neq i}}
|A_{ik}|^2|A_{nk}|^2
J_{ki}
\Big[
-p_kS'(\omega_{kn})
+
p_nS'(\omega_{nk})
\Big]
\nonumber\\
&=
\Psi_{ni}^{r2},
\end{align}

The principal-value contribution of the second representative is
obtained by the transposition
\[
(k n, i k)\leftrightarrow(i k,k n),
\]
and obeys the companion identity
\(
\Psi_{ni}^{p2}
=
\Psi_{ni}^{r1}
\).

Thus, the principal-value contribution of each
representative is mapped by KMS symmetry onto the
resonant contribution of its transposed partner.

Inserting Eqs.~\eqref{Eq:Pem1}and~\eqref{Eq:Pem2}  yields 
\begin{align}
\Psi_{ni}&=\Psi_{ni}^{r1}+\Psi_{ni}^{r2}\nonumber\\
&=
-
p_n\sum_{\substack{k\neq n\\k\neq i}}
|A_{ik}|^2|A_{nk}|^2
(J_{nk} S'_{ki}+J_{ki}S'_{nk})
\label{Eq:Pem1a}
\\
&
+
p_i\sum_{\substack{k\neq n\\k\neq i}}
|A_{ik}|^2|A_{nk}|^2
\Big[
J_{kn}S'_{ik}+
J_{ik}S'_{kn}
\Big].
\label{Eq:Pem2a}
\end{align}
Comparing with Eq.~\eqref{Eq:Psi_asym}, we identify
\[
C_{ni}
=
\sum_{\substack{k\neq n\\k\neq i}}
|A_{ik}|^2|A_{nk}|^2
\Big(
J_{kn}S'_{ik}
+
J_{ik}S'_{kn}
\Big),
\]
as a representative of the same stationary-state-preserving class.

Adding the  contribution \(C_{ni}\) to
Eq.~\eqref{Eq:StandardForm} yields the standard
form of the latent-resonant generator. Its off-diagonal
matrix elements are decomposed into the local and interacting generator contributions, e.g., 
\begin{equation}
L_{nn,ii}^{\rm std}=L_{nn,ii}^{\rm qu}+L_{nn,ii}^{\rm int}.
\end{equation}
Here,
\begin{widetext}
\begin{subequations}\label{Eq:StandardFormX}
\begin{align}
L_{nn,ii}^{\rm qu}
&=
2\vert A_{ni}\vert^4
\big[
(J_{in}-J_{ni})S'_{in}
+
(S_{in}-S_{ni})J'_{in}
+
2J_{in}S'_{ni}
\big]
+
2\vert A_{ni}\vert^2(A_{ii}^2-A_{nn}^2)S_0
\label{Eq:SFembedX}
\\
L_{nn,ii}^{\rm int}&=
\sum_{\substack{k\neq n\\ k\neq i}}
\big[
\vert A_{nk}\vert^2
\big(
2\vert A_{ni}\vert^2J_{in}S'_{nk}
-
\vert A_{ik}\vert^2J_{ik}S'_{kn}
+
\vert A_{ik}\vert^2J_{kn}S'_{ik}
\big)
-
i\vert A_{ni}\vert^2\Gamma'_{in}
\big(
\vert A_{ik}\vert^2\Gamma_{ik}
-
\vert A_{nk}\vert^2\Gamma_{nk}
\big)
\big]
+\mathrm{c.c.},
\label{Eq:StandardFormFinalX}
\end{align}
\end{subequations}
\end{widetext}
which is Eq.~\eqref{Eq:StandardForm} in main text.

To establish sufficiency for R2E$_2$, the reduced
latent-resonance generator must satisfy the stationary state condition
\begin{equation}
\sum_{i\neq n}
\Big(
L^{(2)}_{nn,ii}p_i^{(2)}
-
L^{(2)}_{ii,nn}p_n^{(2)}
\Big)
=
-\sum_{i\neq n}
\Big(
L_{nn,ii}^{\rm std}p_i
-
L_{ii,nn}^{\rm std}p_n
\Big),
\label{Eq:R2EstdTrace}
\end{equation}
which we prove next.

The proof relies repeatedly on the KMS relations Eq.~\eqref{Eq:KMS00} and
\begin{align}
J'_{in}
&=
\beta J_{in}
-
e^{\beta\omega_{in}}J'_{ni},
\end{align}
as well as the useful qubit contraction identity,
valid for $n\neq i$,
\begin{align}
&
L_{nn,ii}^{\rm qu}p_i
-
L_{ii,nn}^{\rm qu}p_n
\nonumber\\
&=
2|A_{ni}|^4(p_i+p_n)
\bigl[
J_{in}S'_{ni}
-
J_{ni}S'_{in}
\bigr]
\nonumber\\
&\quad
-2\beta |A_{ni}|^4p_iJ_{in}
(S_{ni}-S_{in})
+2|A_{ni}|^2(A_{ii}^2-A_{nn}^2)S_0.
\label{Eq:quContract}
\end{align}

\subsection{Second-Order Return to Equilibrium in a Qubit}
\label{App:QubitR2E}

For \(N=2\),  the
restricted sums in Eq.~\eqref{Eq:StandardFormFinalX} over intermediate levels
\(k\neq n,i\) vanish. We will denote the remaining qubit levels as $n$ and $i$.
The right-hand side of
Eq.~\eqref{Eq:R2EstdTrace} is evaluated using
Eq.~\eqref{Eq:quContract} together with the KMS
condition. Then, using the normalization $p_i+p_n=1$ as well as $L^{(2)}_{nn,ii}=2\vert A_{in}\vert^2 J_{in}$, one arrives at
\begin{align}
    p_n^{(2)}&=
    \vert A_{ni}\vert ^2\big(p_nS'_{ni}-p_iS'_{in}
    -\beta p_ip_n(S_{ni}-S_{in})\big)
\end{align}
for a traceless qubit coupling,  which is precisely the mean-force Gibbs correction~\cite{cresser2021weak}.

\subsection{Induction Step\label{Sec:induction}}

The induction step is essential because the first
nontrivial interactions between frequency pairs appear
only for \(N>2\). 

Assume that Eq.~\eqref{Eq:R2EstdTrace} holds for an
arbitrary nondegenerate \(N\)-level system. We next show
that the same identity remains valid after adding one
additional level, \(M=N+1\). By induction, this proves
Eq.~\eqref{Eq:R2EstdTrace} for arbitrary finite
nondegenerate systems.

The additional level has energy \(E_M\) and coupling
matrix elements \(A_{Mi}=A_{iM}^\ast\),
\(i=1,\ldots,N\). The Gibbs probabilities in the
\((N+1)\)-level system, denoted by \(p_i'\), are related
to the original probabilities by
\[
p_i'=\frac{p_i}{1+p_M},
\qquad
p_M=\frac{e^{-\beta E_M}}{Z(N)},
\]
where \(Z(N)\) is the partition function of the
\(N\)-level system. Since every term in the
\((N+1)\)-level identities carries the common factor
\((1+p_M)^{-1}\), this factor will be suppressed
throughout. Likewise, we will not distinguish
notationally between \(p_i'\) and \(p_i\) for
\(i=1,\ldots,N\), while \(p_M'=p_M/(1+p_M)\). This
convention significantly simplifies the expressions
below.

For example, the Gibbs correction~\eqref{Eq:mfgscor}
of the \(M\)-level system is simply
\begin{align}
p_n^{(2M)}
&=
p_n^{(2N)}
+\delta p_{nM}^{(2)}.
\label{Eq:mfgscor1}
\end{align}
The first term reproduces the \(N\)-level mean-force
Gibbs correction, while the second is the contribution
from the additional level \(M\).

Similarly, the second-order generator in the
\(M\)-level system is obtained by extending the index
range. The existing off-diagonal matrix elements are
unchanged, so we do not distinguish notationally
between the \(N\)- and \(M\)-level generators:
\begin{equation}
L^{(2)}_{nn,ii}
=
2|A_{ni}|^2J_{in},
\qquad
n,i=1,\ldots,M,
\quad
n\neq i.
\label{Eq:L2M}
\end{equation}

We want to show
\begin{equation}
\sum_{i\neq n}^M
\Big(
L^{(2)}_{nn,ii}p_i^{(2M)}
-
L^{(2)}_{ii,nn}p_n^{(2M)}
\Big)
=
-\sum_{i\neq n}^M
\Big(
L_{nn,ii}^{\rm std,M}p_i
-
L_{ii,nn}^{\rm std,M}p_n
\Big).
\label{Eq:R2EstdTrace1}
\end{equation}
The induction step \(N\to M\) is obtained by
separating the contributions generated by the
additional level ($M$). 

\begin{figure}[t]
\centering
\includegraphics[width=0.8\linewidth]{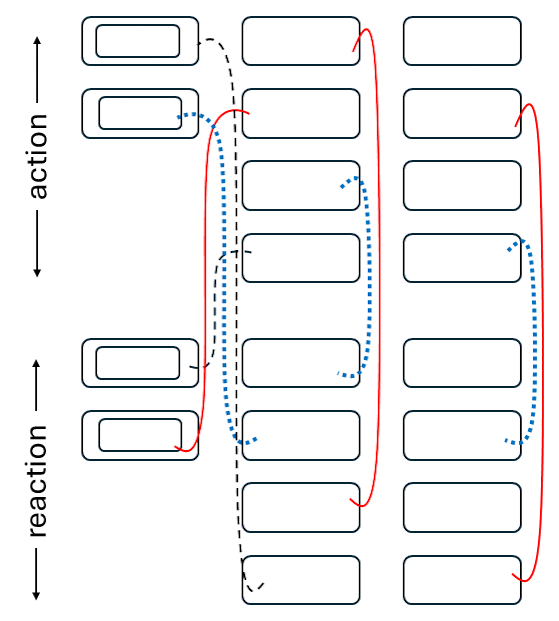}
\caption{
Schematic representation of the cancellations appearing
in the induction step \(N\to N+1\). Each rectangle
represents a contribution to the latent-resonance
interaction shell generated by the addition of the level
\(M=N+1\). The upper block (``action'') contains the
terms induced in the pre-existing \(N\)-level generator
and represents the ten terms in Eq.~\eqref{Eq:LactionI}.
The lower block (``reaction'') contains the contributions
from the new matrix element \(L_{nn,MM}\) and represents
the ten terms in Eq.~\eqref{Eq:LReactionI}. Double boxes
denote terms carrying an additional prefactor of \(2\).
Colored connections indicate direct and KMS-induced
cancellations. After these cancellations, only the newly
generated mean-force Gibbs corrections remain, reproducing the
induction increment.
}
\label{Fig:canceling}
\end{figure}

We begin with the term on the right-hand side of
Eq.~\eqref{Eq:R2EstdTrace1}. Separating the
\(\sum_{i\neq n}^N\) contribution yields the action of
the additional level \(M\) on the pre-existing
\(N\)-level generator:
\begin{widetext}
\begin{align}
&\sum_{i\neq n}^N
\big(
L_{nn,ii}^{\rm std,M}p_i
-
L_{ii,nn}^{\rm std,M}p_n
\big)
=
\sum_{i\neq n}^N
\big(
L_{nn,ii}^{\rm std,N}p_i
-
L_{ii,nn}^{\rm std,N}p_n
\big)
\nonumber\\
&\quad
+
\sum_{i\neq n}^N
\Bigg\{
|A_{nM}|^2p_i
\Big(
2|A_{ni}|^2J_{in}S'_{nM}
-|A_{iM}|^2J_{iM}S'_{Mn}
+|A_{iM}|^2J_{Mn}S'_{iM}
\Big)
\nonumber\\
&\qquad
-p_n|A_{iM}|^2
\Big(
2|A_{ni}|^2J_{ni}S'_{iM}
-|A_{nM}|^2J_{nM}S'_{Mi}
+|A_{nM}|^2J_{Mi}S'_{nM}
\Big)
\nonumber\\
&\qquad
-i p_i|A_{ni}|^2\Gamma'_{in}
\Big(
|A_{iM}|^2\Gamma_{iM}
-
|A_{nM}|^2\Gamma_{nM}
\Big)
\nonumber\\
&\qquad
+i p_n|A_{ni}|^2\Gamma'_{ni}
\Big(
|A_{nM}|^2\Gamma_{nM}
-
|A_{iM}|^2\Gamma_{iM}
\Big)
+\mathrm{c.c.}\Bigg\}
\label{Eq:LactionI}
\end{align}
The first term on the right-hand side isolates the
\(N\)-level contraction, while the restricted sum remainder
contains all additional contributions generated by the
virtual interaction with the added level \(M\).

The complementary reaction arises from the
\(i=M\) term. The additional level \(M\) interacts
directly with \(n\) through the qubit contribution
\(L^{\rm qu}\) and indirectly through the intermediate levels \(i\neq n\): 
\begin{align}
&L_{nn,MM}^{\rm std,M}p_M
-
L_{MM,nn}^{\rm std,M}p_n
=
L_{nn,MM}^{\rm qu}p_M
-
L_{MM,nn}^{\rm qu}p_n
\nonumber\\
&\quad+
\sum_{i\neq n}^N
\Bigg\{
p_M|A_{ni}|^2
\Big(
2|A_{nM}|^2J_{Mn}S'_{ni}
-|A_{Mi}|^2J_{Mi}S'_{in}
+|A_{Mi}|^2J_{in}S'_{Mi}
\Big)
\nonumber\\
&\qquad
-p_n|A_{Mi}|^2
\Big(
2|A_{Mn}|^2J_{nM}S'_{Mi}
-|A_{ni}|^2J_{ni}S'_{iM}
+|A_{ni}|^2J_{iM}S'_{ni}
\Big)
\nonumber\\
&\qquad
-i p_M|A_{nM}|^2\Gamma'_{Mn}
\Big(
|A_{Mi}|^2\Gamma_{Mi}
-
|A_{ni}|^2\Gamma_{ni}
\Big)
\nonumber\\
&\qquad
+i p_n |A_{Mn}|^2\Gamma'_{nM}
\Big(
|A_{ni}|^2\Gamma_{ni}
-
|A_{Mi}|^2\Gamma_{Mi}
\Big)
+\mathrm{c.c.}
\Bigg\}.
\label{Eq:LReactionI}
\end{align}
\end{widetext}

After taking the sum of
Eqs.~\eqref{Eq:LactionI} and~\eqref{Eq:LReactionI},
the interacting contribution undergoes a substantial
reduction. For example, in both the
\(|A_{ni}A_{iM}|^2\) and \(|A_{iM}A_{nM}|^2\) factors,
multiple contributions collapse to only two surviving
terms after a combination of direct and KMS-induced
cancellations. The cancellations arise from repeated
representatives of the same germs and occur exclusively
between the action and reaction equations. 
The pattern of these cancellations is illustrated
schematically in Fig.~\ref{Fig:canceling}.

After the
cancellations, the remaining germs can be fully embedded
into the mean-force Gibbs pairings
\(\delta p_{ab}^{(2)}\) appearing in
Eq.~\eqref{Eq:mfgscor1}, with the final result
\begin{widetext}
\begin{align}
\sum_{i\neq n}^M
\big(
L_{nn,ii}^{\rm std,M}p_i
-
L_{ii,nn}^{\rm std,M}p_n
\big)
&=
\sum_{i\neq n}^N
\big(
L_{nn,ii}^{\rm std,N}p_i
-
L_{ii,nn}^{\rm std,N}p_n\big)+L_{nn,MM}^{\rm qu}p_M-
L_{MM,nn}^{\rm qu}p_n\nonumber\\&+\sum_{i\neq n}^N\Big[-L_{nn,ii}^{(2)}\delta p_{iM}^{(2)}+L_{MM,nn}^{(2)}\delta p_{ni}^{(2)}-L_{nn,MM}^{(2)}\delta p_{Mi}^{(2)}+L_{ii,nn}^{(2)}\delta p_{nM}^{(2)}\Big]
\label{Eq:std2R}
\end{align}

Next, we turn to the left-hand side of
Eq.~\eqref{Eq:R2EstdTrace1}. As before, the
\(M\)-level expression is decomposed into the
\(N\)-level contribution and the increment generated by
the additional level,
\begin{align}
\sum_{i\neq n}^M
\Big(
L^{(2)}_{nn,ii}p_i^{(2M)}
-
L^{(2)}_{ii,nn}p_n^{(2M)}
\Big)
&=\sum_{i\neq n}^N
\Big(
L^{(2)}_{nn,ii}(p_i^{(2N)}+\delta p_{iM}^{(2)})
-
L^{(2)}_{ii,nn}(p_n^{(2N)}+\delta p_{nM}^{(2)})
\Big)\nonumber\\
&+
L^{(2)}_{nn,MM}\sum_{i=1}^M\delta p_{Mi}
-
L^{(2)}_{MM,nn}(p_n^{(2N)}+\delta p_{nM}^{(2)})\nonumber\\
&=\sum_{i\neq n}^N
\Big(
L^{(2)}_{nn,ii}p_i^{(2N)}
-
L^{(2)}_{ii,nn}p_n^{(2N)}
+
L^{(2)}_{nn,ii}\delta p_{iM}^{(2)}
-
L^{(2)}_{ii,nn}\delta p_{nM}^{(2)}
+
L^{(2)}_{nn,MM}\delta p_{Mi}^{(2)}\nonumber\\
&\qquad\qquad
-L^{(2)}_{MM,nn}p_{ni}\Big)+
L^{(2)}_{nn,MM}(\delta p_{MM}^{(2)}+\delta p_{Mn}^{(2)})
-
L^{(2)}_{MM,nn}(\delta p_{nn}^{(2)}+\delta p_{nM}^{(2)})
\label{Eq:actionL2}
\end{align}

Inserting Eqs.~\eqref{Eq:std2R} --\eqref{Eq:actionL2} into Eq.~\eqref{Eq:R2EstdTrace1}, the interaction terms cancel immediately, leaving
\begin{align}
\sum_{i\neq n}^N
\big(
L_{nn,ii}^{\rm std,N}p_i
-
L_{ii,nn}^{\rm std,N}p_n\big)+L_{nn,MM}^{\rm qu}p_M-
L_{MM,nn}^{\rm qu}p_n&=-\sum_{i\neq n}^N
\Big(
L^{(2)}_{nn,ii}p_i^{(2N)}
-
L^{(2)}_{ii,nn}p_n^{(2N)}\Big)\nonumber\\
&-L^{(2)}_{nn,MM}(\delta p_{MM}^{(2)}+\delta p_{Mn}^{(2)})
+
L^{(2)}_{MM,nn}(\delta p_{nn}^{(2)}+\delta p_{nM}^{(2)})
\end{align}
\end{widetext}
The first sum cancels by the induction hypothesis for the
\(N\)-level system. The remaining contribution is 
the normalized R2E$_2$ condition for the
effective qubit \((n,M)\), whose validity was established
previously. Consequently, the increment generated by the
addition of level \(M\) reproduces exactly the mean-force
Gibbs increment, proving Eq.~\eqref{Eq:R2EstdTrace1} for
the \((N+1)\)-level system and completing the induction.

\bibliographystyle{apsrev4-2}
%

\end{document}